\documentclass{aa}  
\usepackage{graphicx}
\usepackage[toc,page]{appendix}
\usepackage{caption}

\usepackage{txfonts}

\usepackage{array}
\newcolumntype{P}[1]{>{\centering\arraybackslash}p{#1}}
\newcolumntype{M}[1]{>{\centering\arraybackslash}m{#1}}

\usepackage{natbib}
\bibpunct{(}{)}{;}{a}{}{,}

\usepackage[dvipsnames]{xcolor}

\usepackage{ulem}

\begin{document} 

   \title{The contribution of faint Lyman-$\alpha$ emitters to extended Lyman-$\alpha$ halos constrained by MUSE clustering measurements}
   \author{Yohana Herrero Alonso\inst{1}, L. Wisotzki\inst{1}, T. Miyaji\inst{2}, J. Schaye
          \inst{3} \and
         J. Pharo\inst{1} \and M. Krumpe\inst{1}
          }
          
   \institute{Leibniz-Institut f\"ur Astrophysik Potsdam (AIP), An der Sternwarte 16, D-14482 Potsdam, Germany\\
              \email{yherreroalonso@aip.de}
         \and
            Universidad Nacional Aut\'onoma de M\'exico, Instituto de Astronom\'ia (IA-UNAM-E), AP 106, Ensenada 22860, BC, M\'exico
         \and
            Leiden Observatory, Leiden University, P.O. Box 9513, 2300 RA, Leiden, The Netherlands
             }

   \date{Received xxx/Accepted xxx} 
 
  \abstract{Detections of extended Lyman-$\alpha$ halos around Ly$\alpha$ emitters (LAEs) have lately been reported on a regular basis, but their origin is still under investigation. Simulation studies predict that the outer regions of the extended halos contain a major contribution from the Ly$\alpha$ emission of faint, individually undetected LAEs. To address this matter from an observational angle, we use  halo occupation distribution (HOD) modeling to reproduce the clustering of a spectroscopic sample of 1265 LAEs at $3<z<5$ from the MUSE-Wide survey. We integrate the Ly$\alpha$ luminosity function to estimate the background surface brightness due to discrete faint LAEs. We then extend the HOD statistics inwards towards small separations and compute the factor by which the measured Ly$\alpha$ surface brightness is enhanced by undetected close physical neighbors. We consider various clustering scenarios for the undetected sources and compare the corresponding radial profiles. This enhancement factor from LAE clustering depends strongly on the spectral bandwidth $\Delta v$ over which the Ly$\alpha$ emission is integrated and can amount to values of $\approx20-40 $ for  small values of $\Delta v$, around $200-400$ km/s, as achieved by recent studies utilizing integral-field spectrographic data. The resulting inferred Ly$\alpha$ surface brightness of faint LAEs ranges between $(0.4-2)\times 10^{20}\;\mathrm{erg}\; \mathrm{s}^{-1}~\mathrm{cm}^{-2}~\mathrm{arcsec}^{-2}$, with a very slow radial decline outwards. Our results suggest that the outer regions of observed Ly$\alpha$ halos ($R \gtrsim 50$~pkpc) could indeed contain a strong component from external (but physically associated) LAEs, possibly even be dominated by them. Only for a relatively shallow faint-end slope of the Ly$\alpha$ luminosity function would this contribution from clustered LAEs become unimportant. We also confirm that the observed emission from the inner regions ($R\le 20-30$~pkpc) is too bright to be significantly affected by clustering. We compare our findings with predicted profiles from simulations and find good overall agreement. We outline possible future measurements to further constrain the impact of discrete undetected LAEs on  observed extended Ly$\alpha$ halos.
 }
  \keywords{large-scale structure -- high-redshift galaxies -- HOD models -- extended Ly$\alpha$ emission }

   \titlerunning{Faint LAE contribution to apparent Ly$\alpha$ SB profiles}
   \authorrunning{Yohana Herrero Alonso et al.}
 
\maketitle
%
%-------------------------------------------------------------------

\section{Introduction}
\label{sec:introduction}
The Ly$\alpha$ line is a paramount cosmological feature for probing early star-forming galaxies, and commonly assists in the detection of high-redshift galaxies, the Ly$\alpha$ emitters (LAEs). The ionizing photons produced by their young stars ionize neutral hydrogen (\ion{H}{i}) atoms in the neighbouring interstellar medium (ISM) and, after recombination, have a two-third probability of being re-emitted as Ly$\alpha$ photons \citep{partridge}. As a result of the complex radiative transfer that the Ly$\alpha$ emission undergoes, a fraction of photons escape the ISM by resonantly scattering with \ion{H}{i} in the circumgalactic and intergalactic media (CGM, IGM). This, among other factors, causes the emission to become diffuse, giving rise to the so-called extended Ly$\alpha$ halos.

Statistically relevant detections of extended Ly$\alpha$ halos (LAHs) employed narrowband imaging observations \citep{hayashino,nilsson,filkenstein}, which restrained the detectable surface brightness (SB) level to $\sim 10^{-18}$~erg~s$^{-1}$ cm$^{-2}$ arcsec$^{-2}$. A significant step forward in terms of limiting SB were the studies by  \cite{steidel10,matsuda,momose14,xue}, who adopted the image stacking approach and extended the SB threshold by an order of magnitude, SB $\sim 10^{-19}$~erg~s$^{-1}$ cm$^{-2}$ arcsec$^{-2}$.  Recently, another major sensitivity improvement was made possible by the Multi-Unit Spectroscopic Explorer (MUSE) instrument at the ESO-VLT \citep{bacon10}, which evened the limiting SB of individual object-by-object measurements to the limits obtained by the stacking of narrowband data. \cite{wisotzki16,leclerq,adelaide,kusakabe22} reported the detection of ubiquitous extended LAHs at $3<z<6$ and found that, on average, the  LAHs detected by MUSE are a factor $4-20$ more extended than their corresponding UV galaxy sizes, presenting median scale lengths of few physical kpc (pkpc). Combining the added depth of MUSE and the signal gain through stacking, \cite{wisotzki18} ascertained extended LAHs at much larger scales ($\approx$ 60~pkpc at $z=3$) than previous studies at similar redshifts ($\approx$ 30~pkpc). Recent works with Hobby-Eberly
Telescope Dark Energy Experiment (HETDEX; \citealt{hetdex}) LAEs at $1.9<z<3.5$ and Subaru LAEs at $z=2.2-2.3$ again roughly doubled the radii over which LAHs can be detected \citep[160~pkpc and 200~pkpc, respectively;][]{hetdex22,zhang23}.

Understanding the characteristics and, in particular, the nature of these extended LAHs provides information as to the spatial distribution and kinematic properties of the CGM \citep{zheng11a} and, more fundamentally, to the processes of formation and evolution of galaxies \citep{bahcall}. The main mechanisms that are believed to contribute to the existence of LAHs are (i) resonant scattering of Ly$\alpha$ photons produced in ionized H{\sc ii} regions of the ISM, (ii) "in situ" recombination, %i.e., local (in the CGM) recombination of H atoms previously ionized, 
(iii) fluorescence by photons from the metagalactic UV background, %(i.e., distant star-forming galaxies and AGN)
and (iv) collisional excitation from cooling gas accreted onto galaxies, also denoted as "gravitational cooling". Processes (i) and (ii) are consequences of local star formation through Lyman continuum radiation emitted from young and massive stars in star-forming galaxies; processes (iii) and (iv) are driven by external influences.

Comparisons between observational constraints and simulation studies have not yet delivered unique conclusions about the dominant origin of LAHs. Furthermore, while some simulation studies \citep[e.g.,][]{DijkstraKramer,gronke} were able to fully explain the observed extended Ly$\alpha$ emission with the processes mentioned above, others \citep{shimizu,lake,ribas16,ribas,mitchell,byrohl} argued that in addition to these factors there is a significant contribution from Ly$\alpha$ emission originating in faint satellite galaxies. This emission would be significant only by its collective effects, as most of these "satellite LAEs" are too faint to be detected individually at the sensitivity of current observations. \cite{lake} assessed that in their simulations the cooling of gas accreted onto galaxies as well as nebular radiation from satellites are the major contributors in the outer halo ($R>2$~pkpc). \cite{mitchell} found that scattering is the dominant mechanism in the inner regions of the SB profile ($R<7$~pkpc, SB of few $\rm{SB}\sim10^{-19}$~erg~s$^{-1}$ cm$^{-2}$), scattering and satellites contribute equally at intermediate scales ($R\approx10$~pkpc, $\rm{SB}\sim10^{-19}$~erg~s$^{-1}$ cm$^{-2}$), and satellites dominate the large scales ($R>20$~pkpc, $\rm{SB}\sim10^{-20}$~erg~s$^{-1}$ cm$^{-2}$ arcsec$^{-2}$). Similar conclusions were derived in \cite{byrohl}, who identified scattering as the major source of SB and held photons originating in dark matter halos (DMHs) in the vicinity of the central galaxy responsible for the flattening of the observed SB profiles at $R> 100$~pkpc. 

%it is known that simulation overpredict the number of satellites. this might indicate that the feedback mechanisms acting in group-and cluster-scale halos appear to be less efficient in quenching the mass assembly of satellites, and/or that quenching occurs much later in the simulations. \url{https://arxiv.org/pdf/2203.10895.pdf}
The existence of faint sources around more massive and brighter galaxies is certainly predicted by the hierarchical Cold Dark Matter (CDM) model of structure formation. Because this is closely related to non-linear clustering, \cite{ribas} considered several clustering scenarios, coupled to the analytic formalism described in \cite{ribas16}, to investigate the plausibility of faint sources generating LAHs. In line with the previously mentioned simulations, they found that faint LAEs are the major contributors to the Ly$\alpha$ SB profiles already at $R>4$~pkpc. In contrast, \cite{kakiichi} applied a model based on galaxy-Ly$\alpha$ forest clustering data to build Ly$\alpha$ SB profiles and could match  a multitude of observed Ly$\alpha$ SB profiles even without considering the contribution from faint LAEs.

Despite the theoretical efforts to distinguish between the numerous contributions to the extended LAHs, observational tests and evidence for the "faint LAE" scenario are scant, mainly because of the extremely low SB of the Ly$\alpha$ emission at large distances from galaxies. \cite{bacon21} detected very extended Ly$\alpha$ emission around LAE overdensities in the MUSE extremely deep field (MXDF), reaching far beyond individual LAHs, but from comparing their data with the semi-analytic model GALICS \citep{garel15} they concluded that much (and possibly all) of the apparently diffuse emission can be accounted for by the collective contribution of discrete faint LAEs surrounding more luminous (detected) ones. 

Still missing is an assessment of the magnitude of this contribution based on observations rather than on theoretical predictions. In this paper, we address this issue by employing LAE clustering statistics in combination with halo occupation distribution (HOD) models to obtain observational constraints on the contribution of the faint LAEs to the extended LAHs. We build on our previous study (\citealt{yohana2}, hereafter HA23), where we used the HOD framework to interpret the clustering of three MUSE LAE samples at $3<z<6$. Each dataset had a different exposure time and thus covered a distinct range of Ly$\alpha$ luminosities within $40.15<\log(L_{\rm{Ly}\alpha}/[\rm{erg}\;\rm{s}^{-1}])<43.35$. We found a strong (8$\sigma$ significance) clustering dependence on $L_{\rm{Ly}\alpha}$, where more luminous LAEs cluster more strongly and reside in more massive DMHs than faint LAEs. 

The paper is structured as follows. In Sect.~\ref{sec:data} we briefly describe the data used for this work. In Sect.~\ref{sec:clustering} we summarize the clustering properties of our galaxy sample. We extrapolate the clustering features to estimate the contribution of undetected LAEs to the extended Ly$\alpha$ halos in Sect.~\ref{sec:cgm}, where we also compare our estimations to recent observational and simulated results. We give our conclusions in Sect.~\ref{sec:conclusions}. 

Throughout the paper, comoving and physical distances are  given in units of $h^{-1}$Mpc and pkpc, respectively, where $h = H_0/100=0.70$. We use a $\Lambda$CDM cosmology and adopt $\Omega_M =$ 0.3, $\Omega_{\Lambda} =$ 0.7, and $\sigma_8 =$ 0.8 \citep{constants}. All uncertainties represent 1$\sigma$ (68.3\%) confidence intervals. 

%-------------------------------------------------------------------

\section{Data}
\label{sec:data}

This paper is based on the results from two different spectroscopic MUSE samples. While we use a sample of LAEs from the MUSE-Wide survey for the clustering constraints, we employ the data from two MUSE deep fields for the comparison to observed Ly$\alpha$ halos (see Sect.~\ref{sec:contribution}). 

\subsection{The MUSE-Wide survey}
The main input to the HOD model of Sect.~\ref{sec:clustering} is based on the clustering constraints of a subset of the LAE MUSE-Wide survey \citep{herenz17,urrutia19}, similar to the LAE dataset used in \cite{yohana2}. The sample is constructed from 100 fields, each spanning 1 arcmin$^2$, observed with an exposure time of one hour, and covering regions of CANDELS/GOODS-S, CANDELS/COSMOS and the Hubble Ultra Deep Field (HUDF) parallel fields. The survey also incorporates shallow (1.6 hours)
subsets of MUSE-Deep data \citep{bacon17,bacon22} within the HUDF in the CANDELS/GOODS-S region. We refer to \cite{urrutia19} for further details on survey build up, reduction and flux calibration of the MUSE data cubes.

Unlike in HA23, we now focus on LAEs with redshifts within $3<z<5$ (and use the same redshift interval for the comparison to Ly$\alpha$ SB profile measurements in Sect.~\ref{sec:contribution}). Because Ly$\alpha$ peak redshifts are typically offset by up to several hundreds of km$\:{\rm s}^{-1}$ from systemic ones (e.g., \citealt{hashimoto15,sowgat,kasper}), we correct the Ly$\alpha$ redshifts following the relations given in \cite{verhamme18}, which assure an accuracy of $\le100$ km$\:{\rm s}^{-1}$ \citep{kasper}. We refer to \cite{yohana}, hereafter HA21, and references therein for further details on the LAE sample construction.

\begin{figure}[tb]
\centering
\includegraphics[width=\columnwidth]{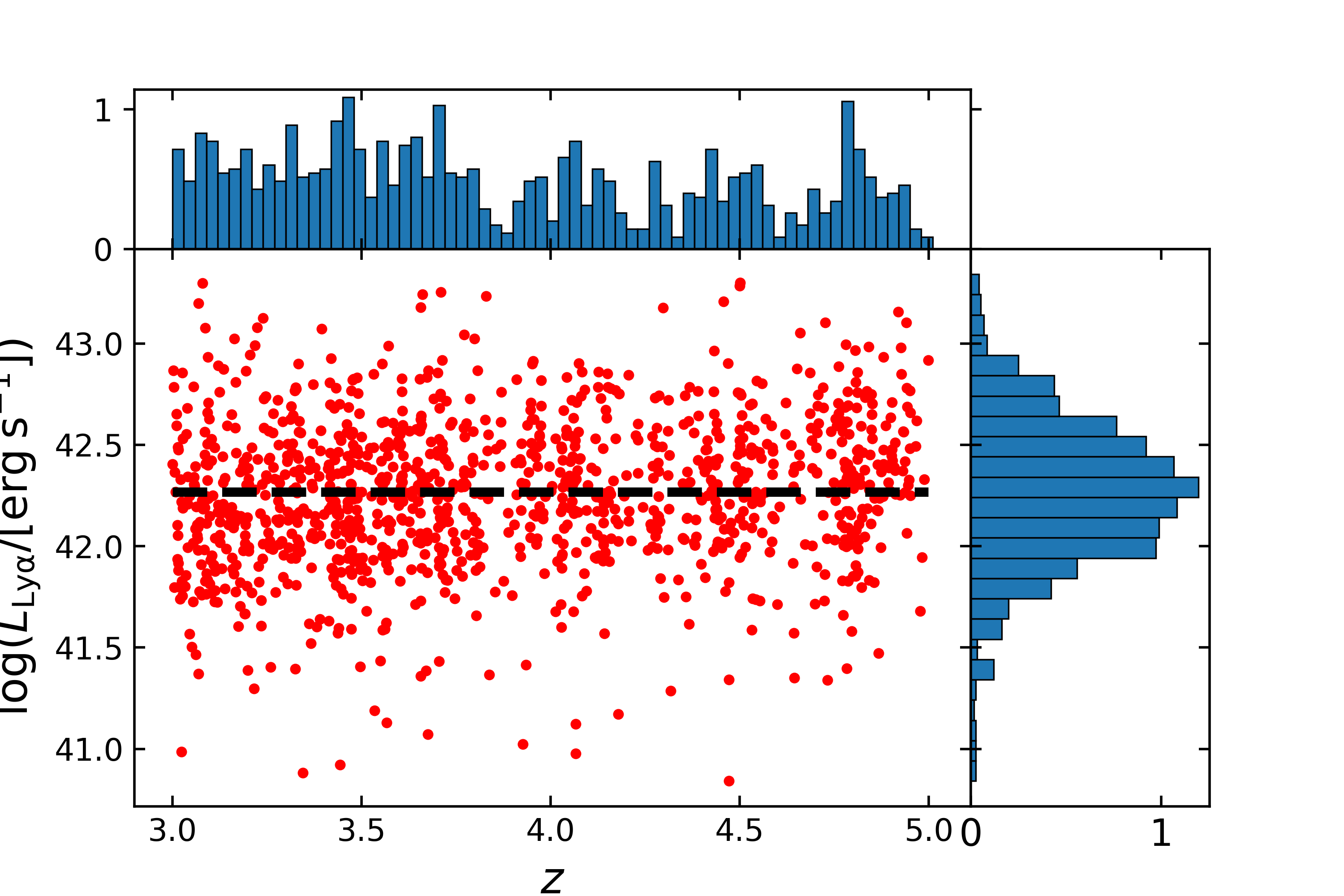}
\caption{Distribution in Ly$\alpha$ luminosity-redshift space of the  $3<z<5$  LAEs selected from the spectroscopic MUSE-Wide survey.
The dashed line corresponds to the median $\log L_{\rm{Ly}\alpha}$ of the sample. The normalized redshift and $\log L_{\rm{Ly}\alpha}$ distributions are shown in the top and right panels, respectively. }
\label{fig:z-LLya}
\end{figure}

Our sample comprises a total of 1265 LAEs with Ly$\alpha$ luminosities in the range 40.84 $<$ log(\textit{L}$_{\rm{Ly}\alpha}/[\rm{erg\:s}^{-1}])$ $<$ 43.30 and a median value of $ \langle\log(L_{\rm{Ly}\alpha}/[\rm{erg\: s}^{-1}])\rangle =42.27$. While the Ly$\alpha$ luminosity and redshift distributions of the sample are presented in Fig.~\ref{fig:z-LLya}, Figs. 1 and B.1 of HA23 show the spatial coverage of the LAE dataset. Taking into account the field-to-field overlaps, the actual surveyed area corresponds to 90.2 arcmin$^2$ and extends $\approx43$ $h^{-1}$Mpc at the median redshift of the sample $\langle z\rangle=3.8$. This implies a LAE density of $\simeq 2\cdot10^{-3}\;h^3\rm{Mpc}^{-3}$. 

%Although there are deeper MUSE surveys and thus less luminous LAE samples available, MUSE-Wide is still the largest of all of them, delivering better statistics and allowing us to have a robust control on the clustering uncertainties. The MUSE-Wide survey is the dataset with the most luminous LAEs and the lowest satellite fraction of all MUSE surveys (see next section).

\subsection{MUSE Deep fields}

In Sect.~\ref{sec:contribution} we compare our results to Ly$\alpha$ radial SB profiles, using the same data and measured in a similar way as by \cite{wisotzki18}. The parent sample is a combination of LAEs detected by MUSE in the Hubble Deep Field South \citep{bacon14} and in the Hubble Ultra-Deep Field \citep{bacon17}, and the definition of the final sample proceeds as in \cite{wisotzki18} except that now we consider only a single set of objects with redshifts $3<z<5$.

To further facilitate comparison to the clustering predictions, we also modify the construction of the resulting Ly$\alpha$ SB profile as follows: Instead of the individually optimized spectral extraction windows for each LAE used by \cite{wisotzki18}, we adopted a fixed velocity bandwidth of 600~km~s$^{-1}$ centered on the peak of the Ly$\alpha$ emission of each galaxy. 
This bandwidth not only
provides a good compromise between noise suppression and flux
loss avoidance in the stacking approach,  it also facilitates the
redshift-space distortion modeling of Sect.~\ref{sec:clustering}. We denote the individually extracted Ly$\alpha$ images as "pseudo-narrowband" in order to distinguish then from genuine filter-based narrowband (NB) imaging data, which encompass much wider bandwidths of typically $\approx12000-25000$~km~s$^{-1}$.
Unlike \cite{wisotzki18}, we also skipped the artificial truncation of the used pseudo-NB data at a radius of $6"$ ($\approx40$~kpc), and we now used the mean instead of the median to stack the images, to make the resulting profiles as similar as possible to the clustering calculations in their construction logic.

%%%%%%%%%%%%%%%%%%%%%%%%%%%%%%%%%%%%%%%%%%%%%%%%%%%%%%%%%%%%%%%%%%%%%%%%%%%%%%%%%%%%%%%%%%%%%%%%%%%%%%%%%%%%

\section{Clustering framework}
\label{sec:clustering}

In HA23 we measured the clustering of three LAE samples, including a subset of the MUSE-Wide survey, with the K-estimator of \cite{adelberger}. The K-estimator measures the radial clustering along line of sight distance, $Z $, by counting galaxy pairs  in redshift space at fixed transverse separations, $R $. %The estimator $K^{0,7}_{7,45}(R)$ quantifies the ratio of galaxy pairs within $0<|Z /[h^{-1}{\rm cMpc}]|<7$ and $0<|Z /[h^{-1}{\rm cMpc}]|<45$ for individual transverse bins, representing thus the excess of galaxy pairs in the first $Z $ bin with respect to the total one (see Eq. 1 in \citealt{yohana}). 
The  K-estimator is directly related to the average underlying correlation function (see Eq. 2 in HA21).  We refer to Sect.~3.1 in HA21 for further details.

We then fitted the K-estimator measurements with HOD modeling performed at the median redshift of the  galaxy pairs of the sample. The HOD model we used is a simplified version of the five parameter model by \citet{zheng07}. Because of sample size limitations, the halo mass at which the satellite occupation becomes zero  and the scatter in the central halo occupation lower mass cutoff were fixed to $M_0=0$ and $\sigma_{\log M}=0$, respectively. The three free parameters are then the minimum halo mass required to host a central galaxy, $M_{\rm{min}}$, the halo mass threshold to host (on average) one satellite galaxy, $M_1$, and the high-mass power-law slope of the satellite galaxy mean occupation function, $\alpha$ (see Sect.~3.3 in HA23). 
Although we included redshift space distortions (RSDs) in the two-halo term using linear theory \citep[Kaiser infall,][]{kaiser,vandenbosch13}, we ignored the effect of RSDs in the one-halo term given its negligible effect at the scales there considered. 

The most relevant finding of HA23 for this work is the strong  clustering dependence on Ly$\alpha$ luminosity. Luminous ($\log(L_{\rm{Ly}\alpha}/[\rm{erg\: s}^{-1}]) \approx42.53$) LAEs cluster significantly ($8\sigma$) more strongly and reside in $\approx25$ times more massive DMHs than less luminous ($ \log(L_{\rm{Ly}\alpha}/[\rm{erg\: s}^{-1}]) \approx40.97$) LAEs. Hence, any fainter dataset is assumed to be less strongly clustered  than those considered here or in HA23.

\begin{figure*}[tb]
\centering
\begin{tabular}{c c}
  \centering
  \includegraphics[width=.47\linewidth]{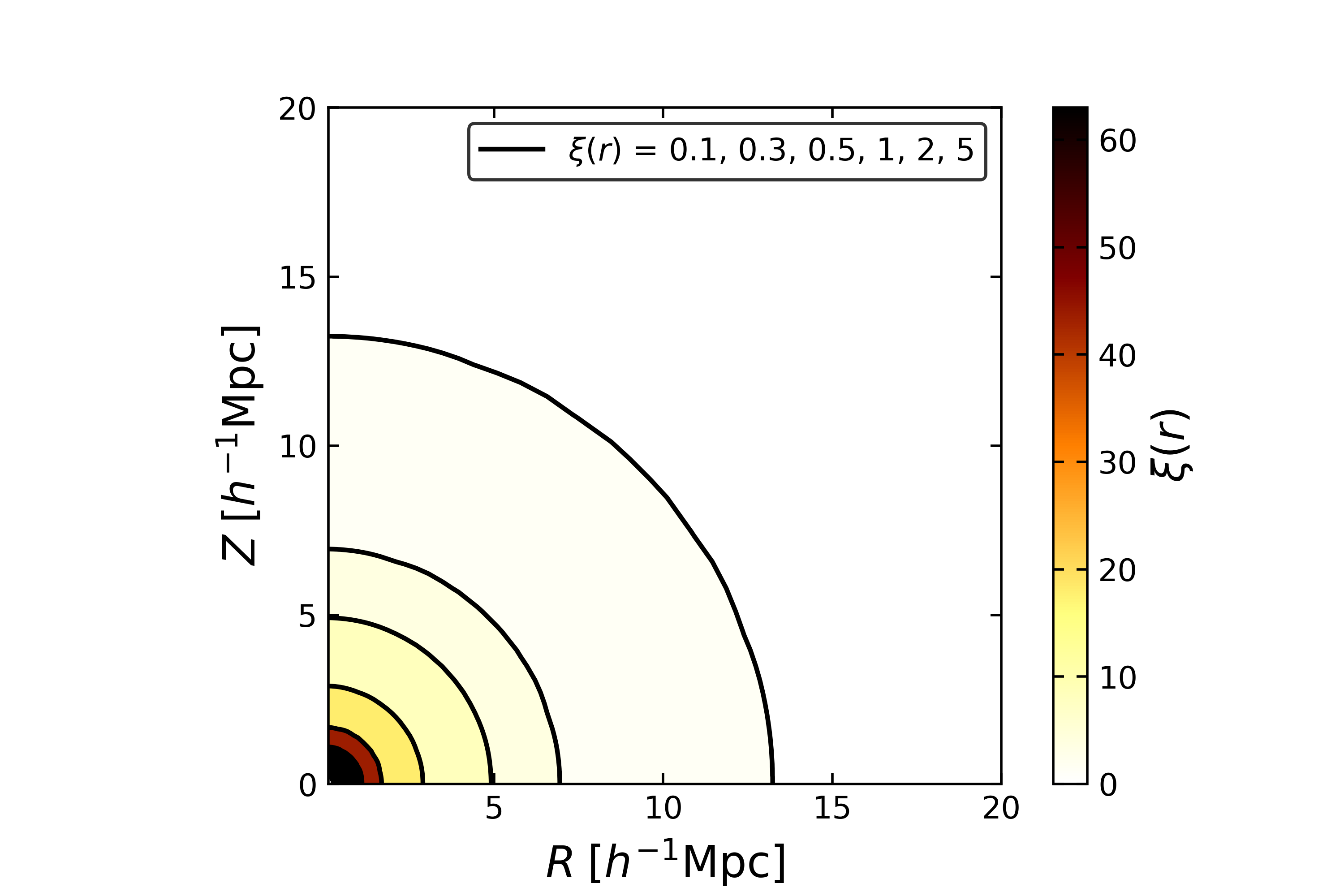}
\end{tabular}
\begin{tabular}{c c}
  \centering
  \includegraphics[width=.47\linewidth]{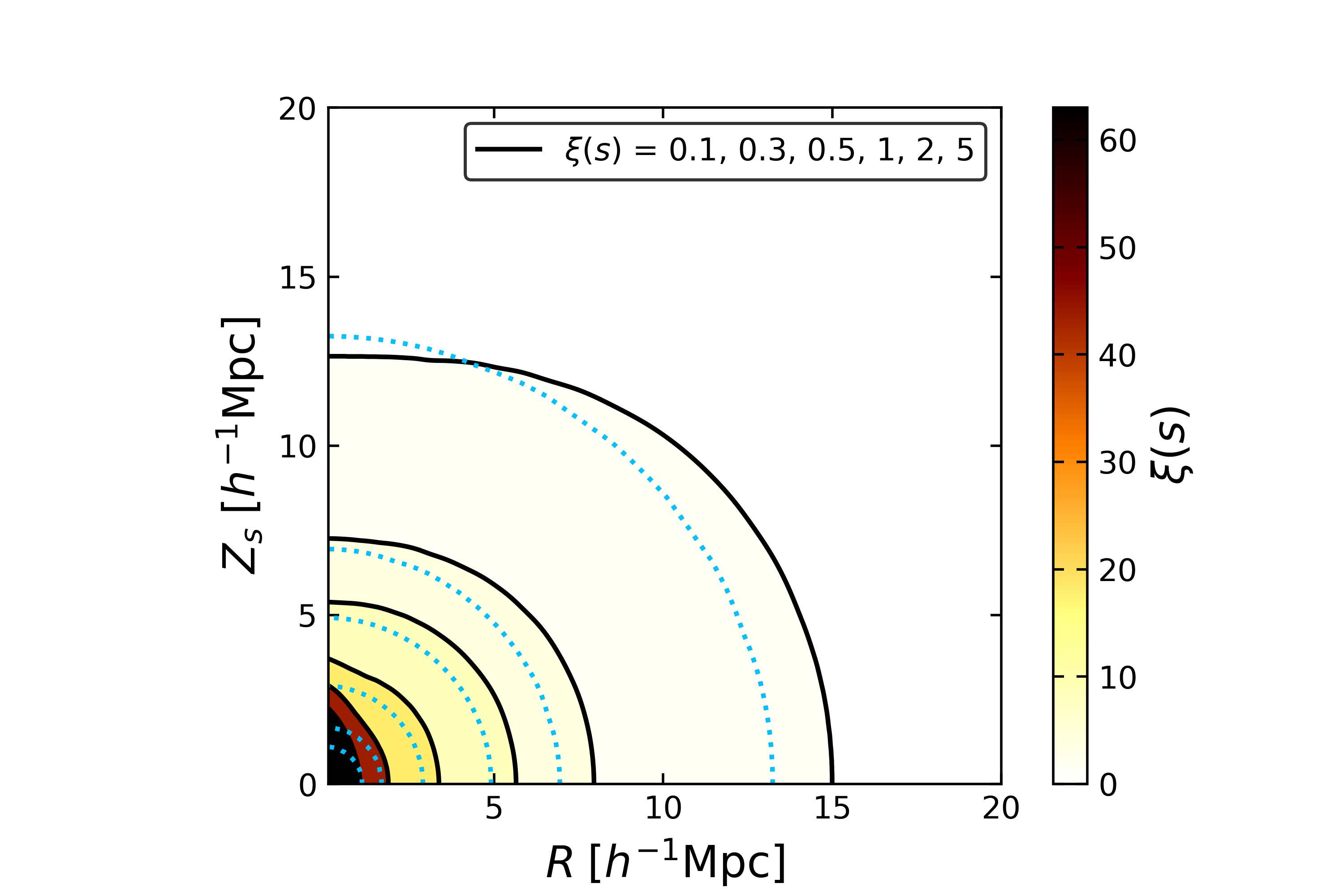}
\end{tabular}
\caption{Best-fit HOD modeled real- and redshift-space 2pcf. Left: Real-space 2pcf, $\xi(r)$, for the sample of MUSE-Wide LAEs at $3<z<5$. Note the circular symmetry of the color-coded contours. Right: Redshift-space 2pcf, $\xi(s)$. The dotted blue contours show the real-space 2pcf from the left panel. Note the elongation of the 2pcf contours along the line-of-sight direction, $Z_s$, at small scales (FoG effect) and the flattening at larger transverse separations, $R$, (Kaiser infall). The contour levels are indicated in the legend.} 
\label{fig:2pcf}
\end{figure*}

In Appendix~\ref{app:same_clustering}, we demonstrate that the clustering strengths of the MUSE-Wide LAE subset considered in HA23 and  our current sample are nearly identical. 
We convert the best-fit HOD model of the K-estimator  found in HA23
to the traditional real-space two-point correlation function (2pcf) using Eq.2 in HA21. %As 2pcf uncertainties we appraise the upper and lower limit of the HOD modeled 2pcfs that fall within the $1\sigma$ probability contours of Fig. 6 in HA23. 
In the left panel of Fig.~\ref{fig:2pcf}
we represent the real-space 2pcf  i.e., $\xi(r)$, where $r=\sqrt{R^2+Z^2}$, whose contours show the expected circular symmetry.  The model parameters correspond to a minimum DMH mass for central LAEs of $\log(M_{\rm{min}} / [h^{-1}\rm{M}_{\odot}])=10.7$, a threshold DMH mass for satellite LAEs of $\log(M_1 / [h^{-1}\rm{M}_{\odot}])=12.4$ and a power-law slope of the number of satellites $\alpha=2.8$. 
The corresponding satellite fraction, typical DMH mass, and virial radius are $f_{\rm{sat}}\approx0.012$, $\log (M_h/[h^{-1}\rm{M}_\odot])=11.09^{+0.10}_{-0.09}$, and $R_{\rm vir}\approx36^{+3}_{-2}$~kpc, respectively (HA23).

Given the small bandwidths of the extracted pseudo-NB Ly$\alpha$ images from MUSE data (a few hundred km~s$^{-1}$ or few comoving Mpc), we now include RSDs in the one-halo term (the so-called Finger-of-God effect, FoG) following \cite{tinker}. We convolve the line of sight component of the real-space 2pcf with the probability distribution function of galaxy pairwise velocities, $P(v)$. For each DMH mass, $M_h$, we assume a Gaussian distribution $P(v,\;M_h)$ with velocity dispersion of central-satellite pairs determined by the streaming model i.e., $\sigma_{\rm v}^2\approx GM_{\rm h}/(2R_{\rm vir})$, where $R_{\rm vir}$ is the virial radius. A Gaussian $P(v,\;M_h)$ is 
supported by hydrodynamic simulations and observational analysis of rich SDSS clusters (see \citealt{tinker} and references therein). We also include the extra contribution corresponding to the uncertainty in the Ly$\alpha$-to-systemic redshift relation, $\sigma_{\rm v, sys}\sim 100$ km s$^{-1}$. However, we ignore the contribution from satellite-satellite pairs given the negligible satellite fraction of the sample. The convolution kernel is then a superposition of weighted Gaussians with dispersion $\sigma_{\rm v}^2\approx GM_{\rm h}/(2R_{\rm vir})+\sigma_{\rm v,sys}^2$.  
We weight the Gaussians with the relative number of satellite galaxies, $N_s$, hosted by each DMH, which involves an integral over the halo mass function $\phi(M_{\rm{h}})\;{\rm d}M_{\rm{h}}$ (see Eq. 5 in \citealt{tinker}) as
\begin{equation}
\label{eq:Pv}
  P(v) =  \frac{\int P(v,\;M_h)\, \langle N_s (M_{\rm{h}})\rangle\, \phi(M_{\rm{h}})\,{\rm d}M_{\rm{h}}}{\int \langle N_s (M_{\rm{h}})\rangle\, \phi(M_{\rm{h}}) \,{\rm d}M_{\rm{h}}}.
\end{equation}

In the right panel of Fig.~\ref{fig:2pcf} we present the redshift-space 2pcf i.e., $\xi(s)=\sqrt{R^2+Z_s^2}$, where $Z_s$ is the line of sight comoving distance in redshift-space. The deviations from circular symmetry are clear. At large scales ($s>5\;h^{-1}$Mpc), the coherent gravitational infall of galaxies onto forming structures \citep[Kaiser infall,][]{kaiser,vandenbosch13} flattens the $\xi(s)$ contours. At small scales ($Z_s<5\;h^{-1}$Mpc),  the stretching of the $\xi(s)$ contours for line of sight separations is caused by the peculiar velocities of galaxies when galaxy redshifts are used as  proxies for distances. Because the distribution of LAEs at $3<z<5$ is affected by the FoG effect only at $Z_s\lesssim5\;h^{-1}$Mpc (peculiar motions of $<500$ km s$^{-1}$) and is negligible at larger separations, we only include the FoG effect when considering velocity widths $<500$ km s$^{-1}$ (see Sect.~\ref{sec:boosting}).

\section{Contribution of faint LAEs to extended Ly$\alpha$ halos}
\label{sec:cgm}

Taking advantage of the clustering constraints, we now seek to assess the contribution of undetected LAEs to the extended Ly$\alpha$ halos. Those objects are much fainter than those considered here or in HA23, and given the overall tendency of galaxies to cluster, they are expected to be found  around our more luminous LAEs. Hence, some fraction of the measured extended Ly$\alpha$ flux must come from those undetected galaxies. While photons included in the central part of the Ly$\alpha$ halo are well distinguished from sky noise, the outskirts of the halo typically have such low SB values that they are close to the noise level. It is on these regions that we focus our attention.

The spatial distribution of these sources enhances the Ly$\alpha$ SB values measured within a given (pseudo-)NB width at any projected distance, $R$, from the observed LAE. This boost is quantified with the clustering enhancement factor, $\zeta(R)$. 
The contribution of these faint LAEs to the apparent Ly$\alpha$ SB profiles also depends on  the number of undetected LAEs, which is obtained from the Ly$\alpha$ luminosity function (LF; $\phi(L){\rm d}L$). The projected Ly$\alpha$ SB profile of undetected LAEs from a given pseudo-narrow band (NB) is:
\begin{equation}
\label{eq:sb}
    {\rm SB}(R)= \zeta(R)\cdot\int_{L_{\rm{min}}}^{L_{\rm{det}}(z)} L\times \phi(L)\:{\rm d}L.
\end{equation}
where $L_{\rm{det}}(z)$ is the detection limit for individual LAEs and $L_{\rm{min}}$ is the adopted lower limit for integrating the LF.

We now discuss the two ingredients of Eq.~\ref{eq:sb} in turn, first the luminosity function and then the enhancement factor.

%%%%%%%%%%%%%%%%%%%%%%%%%%%%%%%%%%%%%%%%%%%%%%%%%%%%%%%%%%%%%%%%%%%%%%%%%%%%%%%%%%%%%%%%%%%%%%%%%%%%%%%%%%%%

\subsection{Ly$\alpha$ luminosity function}
\label{sec:LF}

We quantify the number density of LAEs  as a Schechter function \citep{schechter}:
\begin{equation}
\label{eq:LF}
\phi(L){\rm d}L=\frac{\phi^\star}{L^\star}\; \left(\frac{L}{L^\star}\right)^{\alpha_{\rm LF}} e^{-(L/L^\star)}\;{\rm d}L,
\end{equation}
where $L^\star$ denotes the characteristic luminosity, $\phi^\star$ the normalization density, and $\alpha_{\rm{LF}}$ the faint-end slope of the LF. 

Of greatest importance for our current study is the value of $\alpha_{\rm{LF}}$ because it largely governs the luminosity density ratio of undetected to detected LAEs. While earlier determinations of the Ly$\alpha$ LF could not constrain this parameter very well, this is now improving because of deeper LAE samples, in particular from different MUSE surveys \citep{drake,herenz19,delavieuville}. The emerging trend is that the faint end of the LF is quite steep with $\alpha_{\rm{LF}}$ uncomfortably close to $-2$, which is the limiting value for which the luminosity density would diverge when integrating the LF to $L=0$. But even with slightly less extreme slopes the adopted lower integration limit has a significant impact on the resulting numbers. This probably indicates that the Schechter function approximation breaks down at very low luminosities. Here we simply bypass this uncertainty by considering different lower integration limits as discussed below. 

We also assume that $\alpha_{\rm{LF}}$ is constant within our current redshift range. While \cite{drake} found a tentative indication for a steepening of the faint-end slope towards high $z$, most other studies concluded that the Ly$\alpha$ LF shows little or no significant evolution with redshift for $3 \la z \la 6$ \citep{ouchi08,sobral18,konno18,herenz19,delavieuville}. As our baseline LF prescription we adopt the best-fit parameters of \cite{herenz19}, with $\log(\phi^\star/[{\rm Mpc}^{-3}])=-2.71$, $\log (L^\star/[{\rm erg}\;{\rm s}^{-1}])=42.6$, and $\alpha_{\rm{LF}}=-1.84$.

The upper integration limit in Eq.~\ref{eq:sb} depends  on the details of the specific LAE survey under consideration. %which depends on wavelength due to the varying sensitivity of the MUSE instrument as well as because of the spectral variation of the night sky. 
Since below we compare our calculations with the observed Ly$\alpha$ SB profile constructed from MUSE deep field data, we adopt a flux- and redshift-dependent selection function constructed to characterise this sample. 

For the lower limit we use $\log (L_{\rm{min}}/[\rm{erg}\; \rm{s}^{-1}])=38.5$ as our baseline value, but we also consider higher and lower values. As an extreme lower limit we adopt $\log (L_{\rm{min}}/[\rm{erg}\; \rm{s}^{-1}])\approx37$, which could be generated by the \ion{H}{ii} region around a single O-type star with an  absolute UV magnitude of $M_{\rm{UV}}\approx-6$. More likely candidates for the smallest units that have to be considered are star clusters or dwarf galaxies with several tens or hundreds of such stars. We bracket the expected range by adopting also a "high" value of $\log(L_{\rm{Ly}\alpha}/[\rm{erg}\; \rm{s}^{-1}]) = 40$.

Since our comparison Ly$\alpha$ profile is a mean of LAEs in the redshift range $3<z<5$, we have to average the luminosity density also over this redshift range. In addition to a straight unweighted average, we also calculate a weighted mean that takes the actual redshift distribution of the sample into account. However, these two numbers differ by only  a few percent.

 %This integration delivers a luminosity density, which is then multiplied by the differential comoving volume of the pseudo-NB widths that are used to observationally measure Ly$\alpha$ SB profiles (see Sect.~\ref{sec:boosting}). We convert the differential luminosity density to SB units with the luminosity distance at the median redshift of our sample, $\langle z \rangle=3.8$.

%Because the results from HA23 implied that a survey for relatively luminous LAEs (such as the MUSE-Wide survey) is implicitly biased against low density regions, the latter assumption probably carries a biased Ly$\alpha$ LF. At those faint luminosities, the actual LF  likely presents a steeper slope than $\alpha_{\rm{LF}}=-1.84$, which might be, however, contained within the 1$\sigma$ confidence interval of \cite{herenz19}. 

%Uv lf has a turnover. From bouwens21,22, atek18 it seems that by Muv=-12 there is a turn over meaning that there are only galaxies within Muv=-24 and -10 -> Lya L of 2*10$^44$ to 5*10$^38$ erg/s

\begin{figure}[tb]
\centering
\includegraphics[width=\columnwidth]{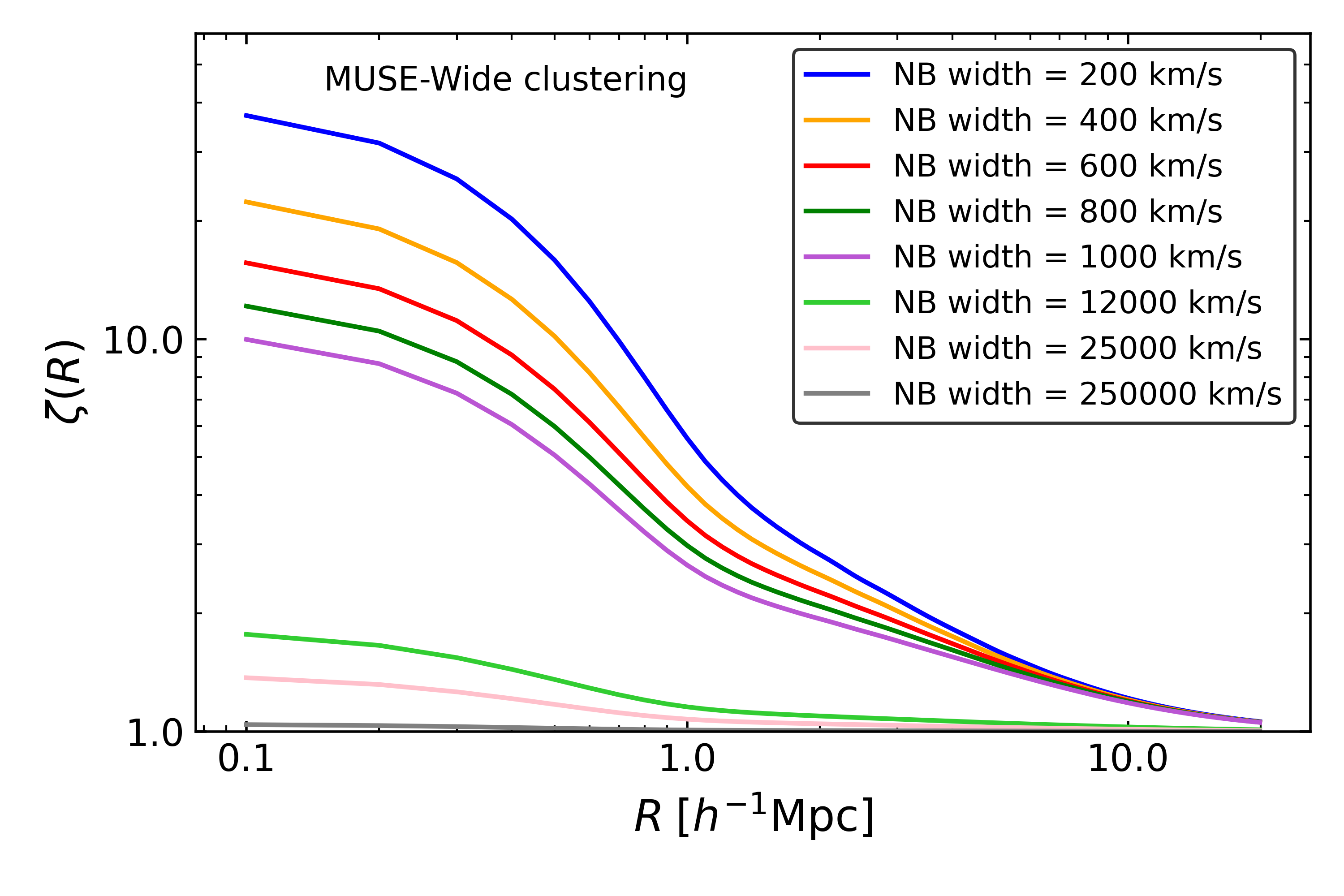}
\caption{Clustering enhancement factor as a function of comoving transverse separation between LAE pairs. Each color corresponds to a different (FWHM) pseudo-NB width. For reference, we include typical NB widths of 12000 and 25000~km~s$^{-1}$ (or 50 and 100~$\AA$) and a broad-band filter width of 250000~km~s$^{-1}$ (or $1000\:\AA$). We assume that both resolved and undetected sources cluster like the MUSE-Wide LAE sample and include RSDs.}
\label{fig:boosting}
\end{figure}

%%%%%%%%%%%%%%%%%%%%%%%%%%%%%%%%%%%%%%%%%%%%%%%%%%%%%%%%%%%%%%%%%%%%%%%%%%%%%%%%%%%%%%%%%%%%%%%%%%%%%%%%%%%%

\subsection{Clustering enhancement factor}
\label{sec:boosting}

\begin{figure}[tb]
\centering
\includegraphics[width=\columnwidth]{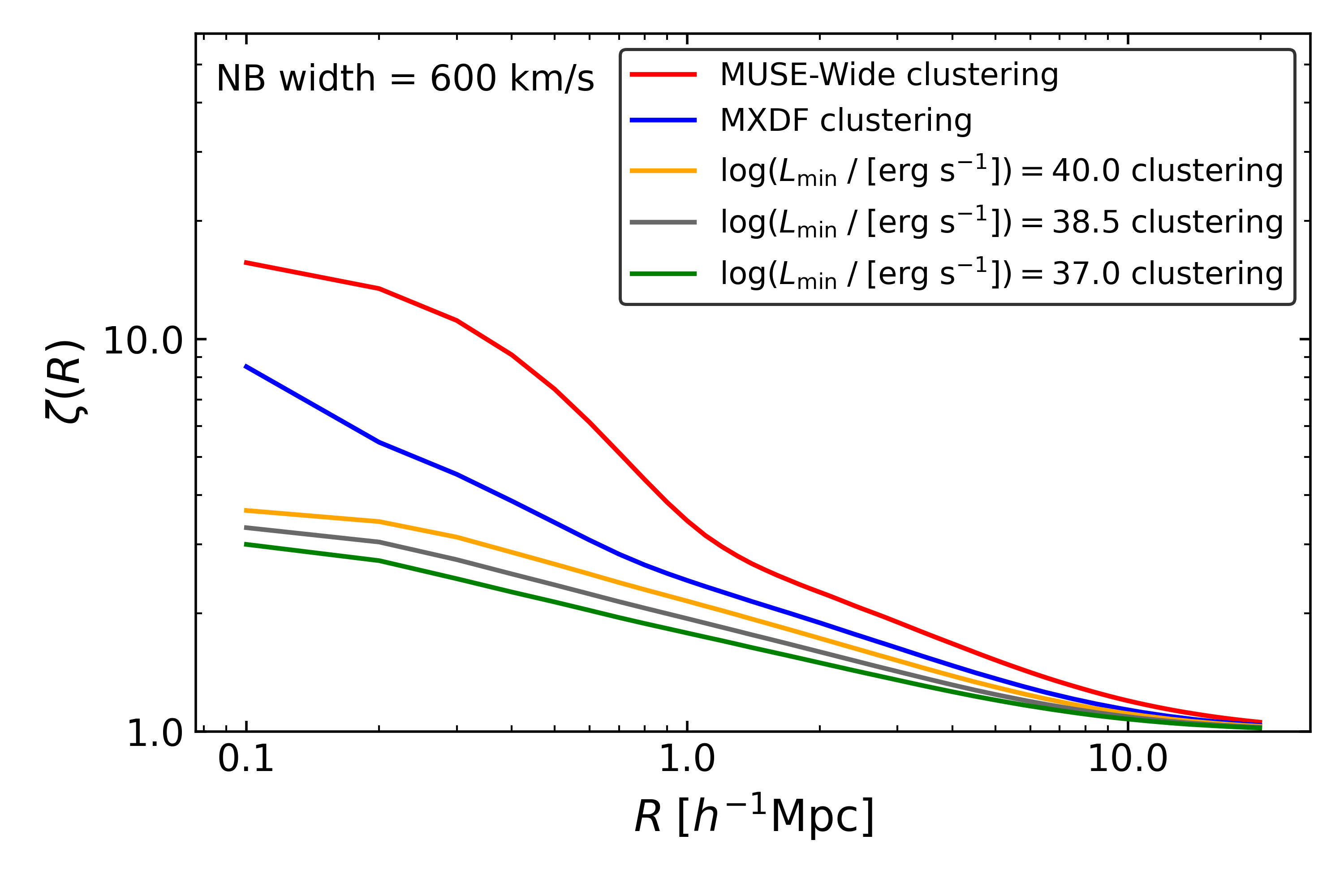}
\caption{As Fig.~\ref{fig:boosting} but for a fixed (FWHM) pseudo-NB width of 600~km~s$^{-1}$ and for different assumed clustering strengths for undetected sources, depending on their Ly$\alpha$ luminosity. The red and blue enhancement factors assume that undetected LAEs cluster like MUSE-Wide and MXDF LAEs, respectively. The orange, gray, and green factors assume an extrapolated HOD 2pcf for LAEs with $\log (L_{\rm{Ly}\alpha}/[{\rm erg \; s}^{-1}])\approx{40.0}$, $\log (L_{\rm{Ly}\alpha}/[{\rm erg \; s}^{-1}])\approx{38.5}$, and $\log (L_{\rm{Ly}\alpha}/[{\rm erg \; s}^{-1}])\approx{37.0}$, respectively. In all cases, detected LAEs cluster like the MUSE-Wide sample.}
\label{fig:boostings_L}
\end{figure}

The enhancement factor represents the boost in Ly$\alpha$ SB due to the spatial distribution of undetected LAEs around our more luminous LAEs. This is computed from the cross-correlation function (CCF) between detected and undetected LAEs, $\xi(R,Z)_{\rm{CCF}}$.  We derive $\zeta(R)$ from the 2pcf  definition \citep[][]{peebles1980} in Appendix~\ref{app:boosting} and display the outcome here: 
\begin{equation}
\label{eq:boosting}
\zeta(R)=\int_{-Z_{{\rm NB}}}^{+Z_{{\rm NB}}} [\xi(R,Z)_{\rm{CCF}}+1]\cdot{\rm d}Z/\Delta Z,
\end{equation}
where $\Delta Z=+Z_{{\rm NB}}-(-Z_{{\rm NB}})$. The radial comoving separations in redshift space,  $Z_{{\rm NB}}$, i.e., including the RSD effect, over which the integral is performed correspond to typical half-widths at half maximum of the (pseudo-)NBs applied in the measurement of Ly$\alpha$ SB profiles ($\sim100-400$~km~s$^{-1}$; see e.g., \citealt{wisotzki18}). In particular, because of the narrowness of $Z_{{\rm NB}}$, it is imperative to model the RSD effects properly to estimate $\xi(R,Z)_{\rm{CCF}}$ from the HOD models. 

Figure~\ref{fig:boosting} shows the variation of the enhancement factor for various  velocity widths (full width at half maximum, FWHM) applied in stacking experiments of (pseudo-)NB images of LAEs. As discussed below, we initially assume that detected and undetected LAEs share the same clustering properties. We thus employ the best-fit HOD modeled 2pcf of the right panel of Fig.~\ref{fig:2pcf}, which corresponds to our sample of $3<z<5$ MUSE-Wide LAEs.
%We cross-correlate the measured clustering of the MUSE-Wide LAEs with the same one assumed for the undetected LAEs. 

\begin{figure*}[tb]
\centering
\includegraphics[width=\textwidth]{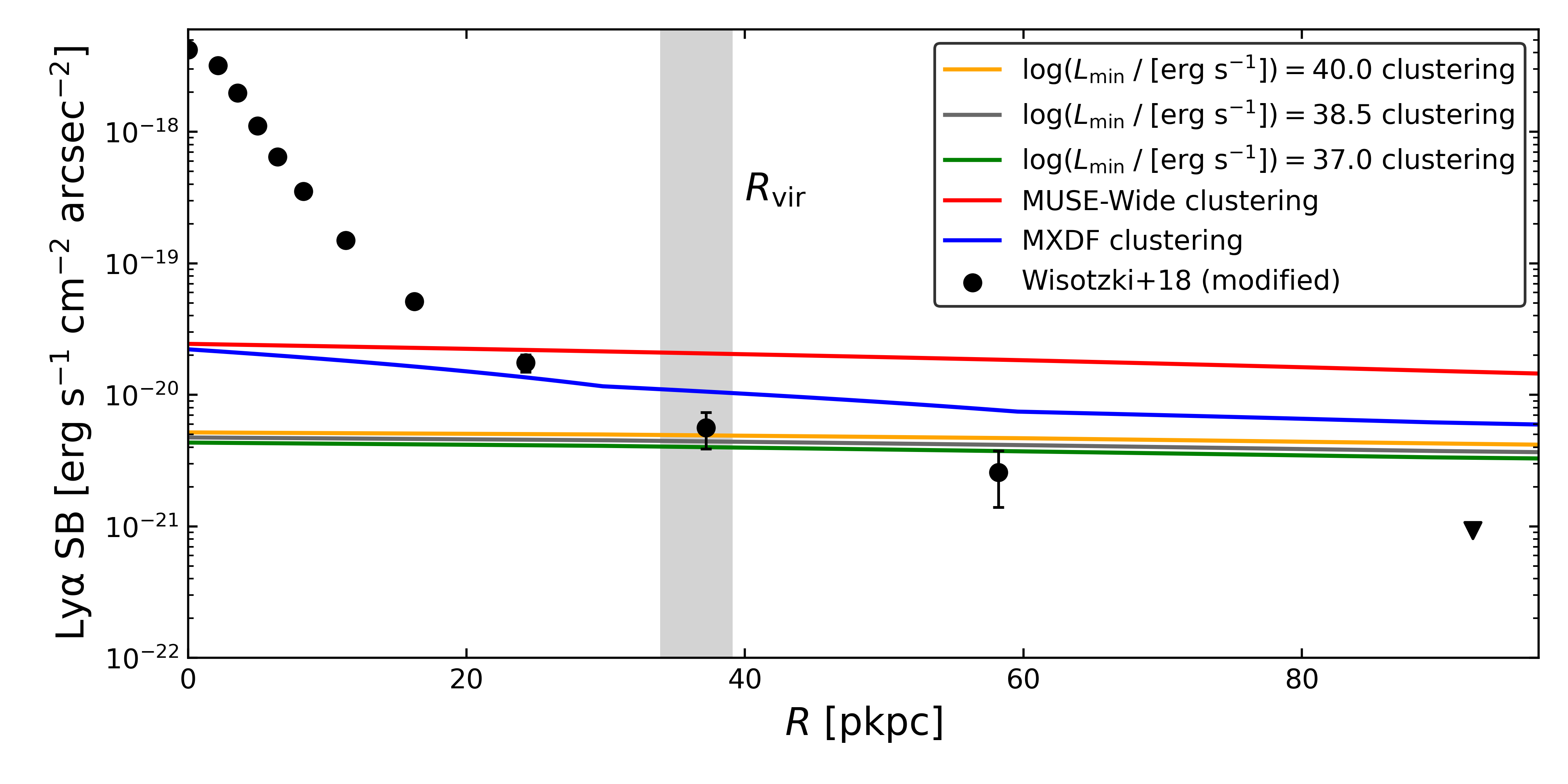}
\caption{Contribution to the Ly$\alpha$ SB profiles from clustered and undetected LAEs with $\log(L_{\rm{min}}/[\rm{erg}\; \rm{s}^{-1}])=38.5$. Different colored lines  correspond to a different  assumption for the clustering of undetected LAEs. Detected galaxies are assumed to cluster like those in the MUSE-Wide sample. The data points correspond to the modified stacking experiment of individual LAEs of \cite{wisotzki18} (see Sect.~\ref{sec:data} and \ref{sec:contribution}). The triangle denotes an upper limit. We employ a pseudo-NB width of 600~km~s$^{-1}$ and the Ly$\alpha$ LF of \cite{herenz19}. The shaded gray region shows the typical virial radius of MUSE-Wide LAEs. } 
\label{fig:SB_clustering}
\end{figure*}

While typical pseudo-NB widths of $200-800$~km~s$^{-1}$ (or $0.5-4\:\AA$) result in substantial enhancement factors of $\approx10-40$ at $R=0.1\;h^{-1}$Mpc, the effective SB values measured with common NB filters FWHM $\approx$ $50-100$~\AA\ or $12000-25000$~km~s$^{-1}$ are enhanced by only $\zeta(R)\approx1.5$--2 and thus barely boosted by clustering. Unsurprisingly, measurements through broad-band filters with FWHM  $\approx$ 1000\:\AA\ (250000~km~s$^{-1}$) remain entirely unaffected ($\zeta(R)\approx1$). Thus, only for the narrow bandwidths achieved by IFU-based Ly$\alpha$ images is there a significant contribution of clustered LAEs to any underlying truly diffuse Ly$\alpha$ emission. %whereas typical NB filters mainly measure mostly the integrated signal ofrom physically unrelated LAEs, even beyond the inner LAH regions.

%The decrease of the enhancement factor with increasing velocity width is due to the larger separations between the LAE pairs measured with broader (pseudo-)filter widths. 

Unless explicitly specified, in the following we  fix the bandwidth to $Z_{\rm{NB}}=3~h^{-1}$Mpc  (intermediate FWHM of $Z=6~h^{-1}$Mpc or 600~km~s$^{-1}$, see Sect.~\ref{sec:data}), corresponding to a window spanning $2.5\:\AA$ in the rest frame around $\lambda_{\rm{Ly}\alpha}\approx1216\:\AA$. 

We also have to make assumptions about the clustering behaviour of the undetected sources to calculate the corresponding enhancement factor. Because the FoG effect is negligible at $Z>5~h^{-1}$Mpc (see Fig.~\ref{fig:2pcf}), in the following we do not include RSDs in the HOD modeling: 

\begin{itemize}
    \item As a first step, we assumed that detected and undetected LAEs present the same clustering properties. We show the corresponding boost factor  in red in Fig.~\ref{fig:boostings_L} (same as the red curve in Fig.~\ref{fig:boosting} but without including RSDs; see the negligible difference between the two curves).  This is however likely to be an overestimate: As demonstrated in HA23, the clustering strength depends significantly on Ly$\alpha$ luminosity (8$\sigma$ in HA23), and therefore the even fainter undetected sources probably cluster less strongly than the  MUSE-Wide LAEs. Since the enhancement factor depends directly on the clustering strength, this option produces only an upper limit on $\zeta(R)$.
    \item We then modify the assumptions and allow the  undetected sources to have a clustering strength similar to the fainter LAEs detected at $3<z<5$ in the MXDF (HA23), which are approximately one order of magnitude less luminous than the MUSE-Wide sample. As shown in Appendix~\ref{app:same_clustering} for MUSE-Wide, the HOD model derived in HA23 for $3<z<6$ can also describe the clustering of the current MXDF dataset. The resulting boost factor is shown in blue in Fig.~\ref{fig:boostings_L}.  Although fainter than MUSE-Wide LAEs, the MXDF LAEs are still considerably more luminous than the hidden undetected ones.  Hence, this boost factor is also an upper limit.
    \item To obtain our best estimate of the actual enhancement factor of the undetected LAEs, we utilize the MUSE-Wide and MXDF HOD clustering probability contours of Sect.~4 of HA23 and extrapolate them to fainter luminosities. We first assume that undetected LAEs  have luminosities of $\log L_{\rm{Ly}\alpha} \approx10^{40}$~erg~s$^{-1}$ and obtain a HOD model with $\log (M_{\rm{min}}/[h^{-1}\rm{M}_{\odot}])=9.9$, $\log (M_1/M_{\rm{min}})=1.1$ and $\alpha=0.1$. We plot the resulting boost factor in orange in Fig.~\ref{fig:boostings_L}. We follow the same procedure for the other two lower integration limits (the resulting HOD models have $\log (M_{\rm{min}}/[h^{-1}\rm{M}_{\odot}])=9.4$, $\log (M_1/M_{\rm{min}})=0.7$, $\alpha=-1.6$, and  $\log (M_{\rm{min}}/[h^{-1}\rm{M}_{\odot}])=8.8$, $\log (M_1/M_{\rm{min}})=0.3$, $\alpha=-3.3$, respectively) and plot their enhancement factors in gray and green.  While the three last enhancement factors are  based on extrapolations and thus currently untestable, we presume that these are probably closer to the truth than the one resulting from the assumption that faint LAEs cluster in the same way as much more luminous objects.
\end{itemize}

%The shapes of the enhancement factor follows the behaviour of the HOD-modeled CCFs. We find markedly weaker enhancements with decreasing line luminosities. While $\zeta(R=0.1\;h^{-1}\rm{Mpc})\approx15$ for undetected sources clustering like those in the MUSE-Wide sample, $\zeta(R=0.1\;h^{-1}\rm{Mpc})\approx3$ for $\log L_{\rm{Ly}\alpha} \approx10^{37.0}$~erg~s$^{-1}$ LAEs clustering like the extrapolated HOD modeled 2pcf assumed for a population of the same luminosity. 

%Because all HOD models are performed at $z=3.8$, we assume that there is no clustering evolution within $3<z<5$. Although neither in HA21 nor in HA23 we found a clustering dependence on redshift, we verify the robustness of this conjecture by studying the clustering of a subsample of our LAEs without the objects at $z>3.8$. The new clustering constraints differ in less than $1\sigma$ from the best-fit HOD model used here and ascertained in HA23.

%We also extrapolate the $L_{\rm{Ly}\alpha}-M_{\rm{h}}$ relation found in Sect.~5.1 of HA23 down to $\log\langle L_{\rm{Ly}\alpha}\rangle\approx10^{38.50}$~erg~s$^{-1}$ (typical DMH mass of $\log (M_{\rm{h}}/[h^{-1}M_{\odot}])\approx8.5$) and estimate an enhancement factor of $\zeta(R)\approx5$.  

 %then sb $5.1\cdot10^{-22}$

To build the Ly$\alpha$ SB profiles, we  have to extrapolate the CCF models down to $R=0$~kpc, since our clustering measurements do not reach the smallest scales of the Ly$\alpha$ SB profiles ($R<30$~pkpc). We also convert the comoving radii to physical kpc.

%The comoving volume of $Z_{\rm{NB}}=3~h^{-1}$Mpc at $\langle z \rangle$ is $8.5\cdot10^{-5}$~Mpc$^3$~arcsec$^{-2}$.

%In the following, whenever we employ an enhancement factor computed from detected LAE populations, our calculations are restricted to be upper limits.

%%%%%%%%%%%%%%%%%%%%%%%%%%%%%%%%%%%%%%%%%%%%%%%%%%%%%%%%%%%%%%%%%%%%%%%%%%%%%%%%%%%%%%%%%%%%%%%%%%%%%%%%%%%%

\subsection{Ly$\alpha$ surface brightness profile from undetected LAEs}
\label{sec:sb}

In Fig.~\ref{fig:SB_clustering} we build the expected Ly$\alpha$ SB profile from undetected LAEs with $\log (L_{\rm{min}}/[\rm{erg}\; \rm{s}^{-1}])=38.5$ for a velocity width of the pseudo-NB of 600~km~s$^{-1}$. The colors show the clustering scenarios considered in Sect.~\ref{sec:boosting}. 
For comparison, we overplot the mean Ly$\alpha$ SB profile based on the stacking of LAEs in the MUSE Deep Fields (Sec.~\ref{sec:data}). We also indicate the typical virial radius of MUSE-Wide LAEs for guidance. 

Before comparing the observed profile with the expected contribution from clustered faint LAEs (see Sect.~\ref{sec:contribution} below) we have to evaluate the uncertainties and dependencies of our calculations with respect to the assumed input parameters. 

We first consider the different adopted clustering scenarios.
By design, the maximum contribution is reached for the MUSE-Wide clustering assumption ($\approx2.5\cdot10^{-20}$~cgs), whereas the lowest extrapolated HOD model delivers SB levels lower by one order of magnitude ($\approx4\cdot10^{-21}$~cgs). Nevertheless, all versions of our estimates exceed the measurements at $R>60$~kpc. 

\begin{figure}[tb]
\centering
\includegraphics[width=\columnwidth]{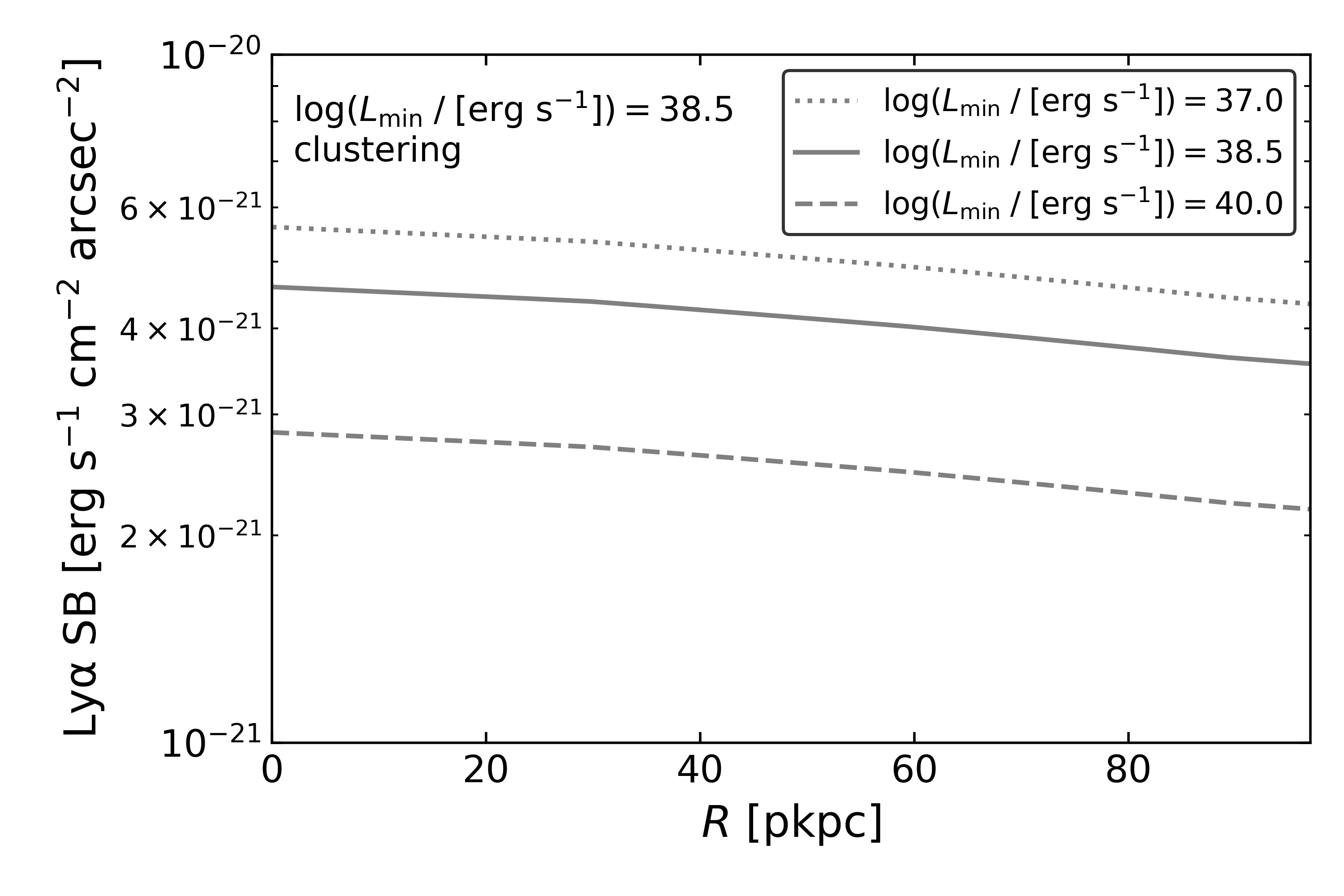}
\caption{Ly$\alpha$ SB variation for plausible luminosities of the undetected LAEs. We assume that undetected sources cluster like the extrapolated HOD model for $\log(L_{\rm{min}}/[\rm{erg}\; \rm{s}^{-1}])=38.5$. Resolved sources cluster like those in the MUSE-Wide sample and the pseudo-NB width is 600~km~s$^{-1}$. We use the  Ly$\alpha$ LF of \cite{herenz19}.} 
\label{fig:SB_Lmin}
\end{figure}

The impact of using different lower integration limits for the LF is evaluated in Fig.~\ref{fig:SB_Lmin}. Here we assume that the clustering of faint LAEs follows the extrapolated HOD model for $\log (L_{\rm{min}}/[\rm{erg}\; \rm{s}^{-1}])=38.5$ LAEs. The three curves show the variation in the SB contribution for various $L_{\rm{min}}$ choices. As expected, the inclusion of fainter sources produces higher Ly$\alpha$ SB levels, although the difference is only a factor of $\sim$2 for the adopted range of $L_{\rm{min}}$.

%In Appendix~\ref{app:boostingKestimator}, we show an alternative and more straightforward method to derive the clustering enhancement factor, directly from the K-estimator. In this case, the enhancement factor is just the ratio of the best HOD fit to the K-estimator measurement and the K-estimator zero-clustering baseline i.e., $7/45$. The theoretical definition of the K-estimator (see Eqs.~2 and 3 in HA21) includes the integration of $\xi(R)$ over the first line of sight separation bin chosen to count the galaxy pairs. In our case, that bin corresponds to $0-7\;h^{-1}$cMpc, in contrast to the $\pm2.5$~cMpc applied in Eq.~\ref{eq:boosting} of the 2pcf approach. Despite the methodological dissimilarities, the two procedures approximately deliver  the same enhancement factor. However, in Sect.~\ref{sec:contribution} we will compare our estimations to the Ly$\alpha$ SB stacking experiment of \cite{wisotzki18}, who apply individually optimized pseudo-NB widths of about 200 km s$^{-1}$. In order to avoid redoing their stacking experiment with NB widths matching the 7~$h^{-1}$cMpc of the K-estimator i.e., 700 km s$^{-1}$, we stick to the 2pcf approach as our main method.

In Fig.~\ref{fig:SBratio} we vary the value of the faint-end LF slope $\alpha_{\rm LF}$ in steps of $\Delta\alpha = \pm 0.1$ and each time recompute the SB profiles. The changes are substantial and demonstrate that the biggest single uncertainty in this estimate is still the faint-end shape of the Ly$\alpha$ LF. 
We also explored modifying the other LF parameters $\phi^*$ and $L^*$, but these have a negligible effect on the predicted Ly$\alpha$ SB profiles.

%Considering the statistically accepted range of faint-end slopes increases the uncertainty on the SB profile in about a factor two. While the clustering contribution becomes more uncertain at smaller scales, the LF error is imprinted in the SB profile equally at all scales. The clustering errors shift the SB values in $\approx5\cdot10^{-21}$~cgs, in contrast to the $\approx10^{-20}$~cgs from the LF.

%The selection of plausible $L_{\rm{det}}$ ($40.5\lesssim\log(L_{\rm{det}}/[\rm{erg}\;\rm{s}^{-1}])\lesssim41.5$) also modifies the SB profile. Appraising ... instead of ..., varies the SB values in ... \yh{give percentages of~cgs values of change} Broadly speaking, fainter luminosity detection thresholds broaden the LF confidence regions of the SB profile because of the more uncertain faint-end slope. 

%%%%%%%%%%%%%%%%%%%%%%%%%%%%%%%%%%%%%%%%%%%%%%%%%%%%%%
%%%%%%%%%%%%%%%%%%%%%%%%%%%%%%%%%%%%%%%%%%%%%%%%%%%%%%

\subsection{Comparison with observed surface brightness profile}
\label{sec:contribution}

It is clear from Fig.~\ref{fig:SB_clustering} that at $R\lesssim 20$~pkpc the stacked Ly$\alpha$ SB is much higher than the level expected from undetected LAEs, irrespective of the clustering scenario, whereas for radii $\ga 50$~pkpc the observed SB level is always comparable to or lower than the contribution estimated from clustering. We note that even where the data points are formally below our calculations e.g., $R=90$~kpc, the combination of error bars in the data with systematic uncertainties in the clustering and luminosity function evaluation imply that these two are not in contradiction.

When interpreting Fig.~\ref{fig:SB_clustering} we have to keep in mind that the two upper (red and blue) curves are upper limits in the sense that for these curves the clustering of the undetected LAEs is assumed to be as strong as for (different sets of) detected LAEs. Allowing for a weaker clustering of the undetected objects shifts the radius of approximate equality outwards to at least $\approx40$~pkpc.

\begin{figure}[tb]
\centering
\includegraphics[width=\columnwidth]{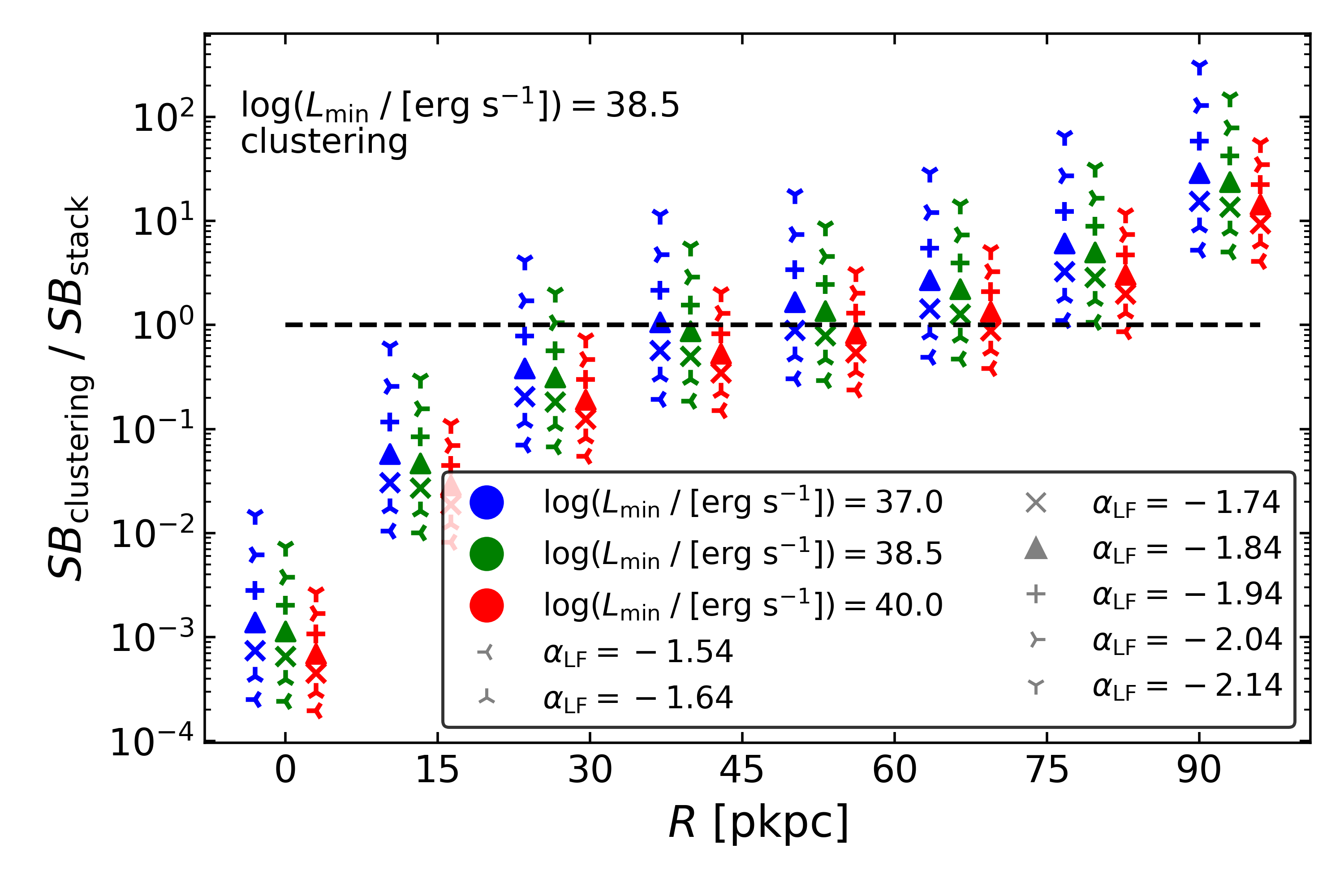}
\caption{SB ratio of the Ly$\alpha$ SB profile from undetected LAEs  and the redone stacked profile from \cite{wisotzki18} at $3<z<5$. The ratios at $R=90$~pkpc are lower limits. Undetected LAEs are assumed to cluster like the extrapolated HOD model for $ L_{\rm{min}}=10^{38.5}$~erg s$^{-1}$ and resolved sources like those in the MUSE-Wide sample. The colors represent different minimum Ly$\alpha$ luminosities for the undetected sources. The symbols display various faint-end slopes of the Ly$\alpha$ LF. The dashed line shows a scenario in which the undetected LAEs alone fully explain the apparent Ly$\alpha$ SB profile. A pseudo-NB width of 600~km~s$^{-1}$ is assumed.}
\label{fig:SBratio}
\end{figure}

Figure~\ref{fig:SBratio} shows the inferred SB ratios (clustering/observed) instead of measured or calculated SB levels. 
For the fiducial baseline faint-end LF slope of $\alpha_{\rm{LF}}=-1.84$, the Ly$\alpha$ emission from undetected LAEs can account for all of the measured Ly$\alpha$ SB profile at $R\gtrsim50$~pkpc, but only a fraction of 20\% at $R\sim 30$~pkpc.
At these smaller distances to the central galaxy we presumably see genuine diffuse emission, powered by the above mentioned  mechanisms.

\begin{figure*}[tb]
\centering
\includegraphics[width=\textwidth]{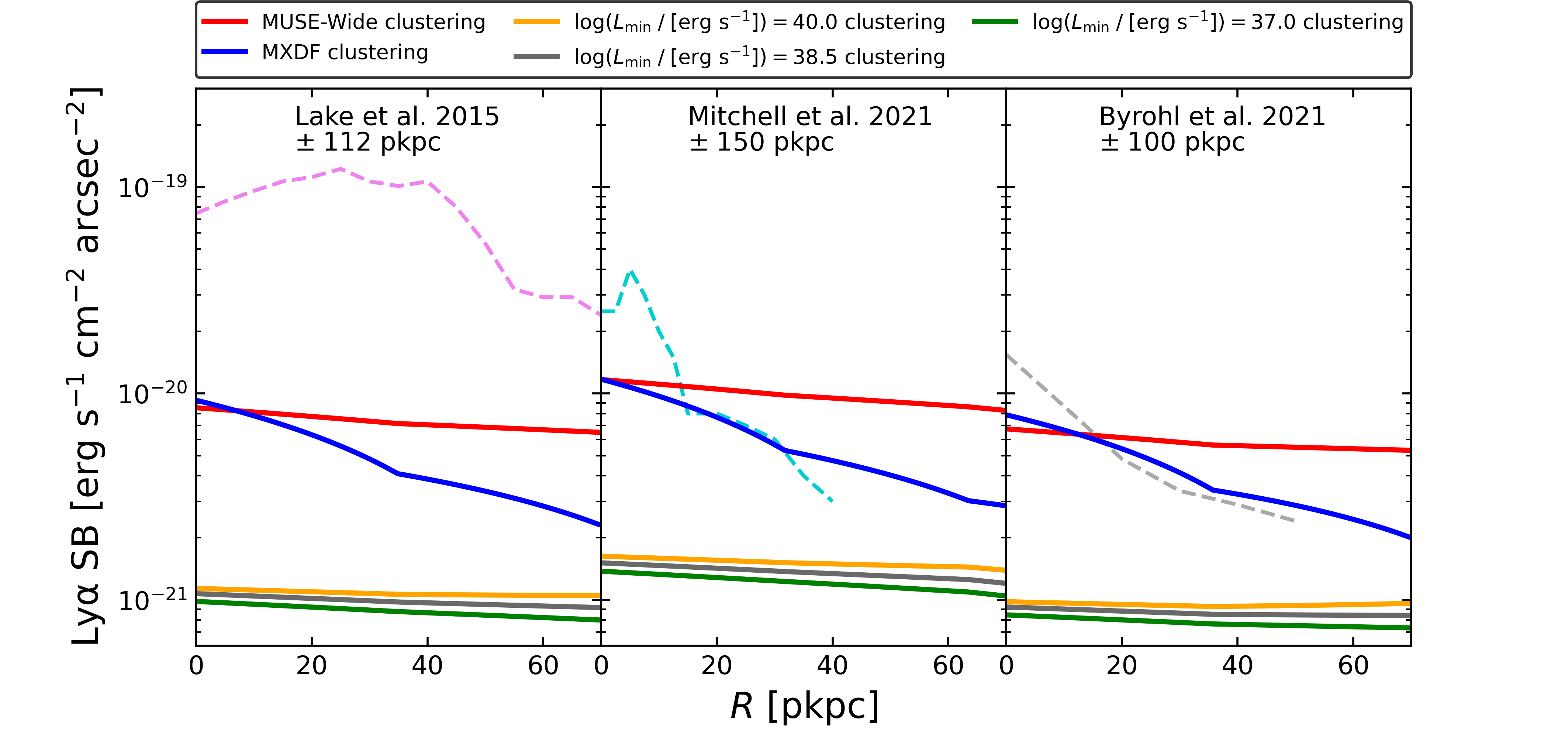}
\caption{Comparison of our Ly$\alpha$ SB profiles from undetected LAEs of $\log(L_{\rm{min}}/[\rm{erg}\; \rm{s}^{-1}])=38.5$ (solid lines)  and the satellite radial profiles predicted from simulations  (dashed lines). The different colors of the solid curves correspond to the different clustering assumptions for the undetected LAEs displayed in the legend. Resolved LAEs are assumed to cluster like those in the MUSE-Wide sample. We have adjusted our pseudo-NB width of Eq.~\ref{eq:boosting} to the projection depth of the simulations.  Left panel: Comparison to L15, whose applied projection depth is equivalent to a pseudo-NB width of 66~km~s$^{-1}$ at $z_{\rm simul}=3.1$. Middle panel: Comparison to M21 with a pseudo-NB width of 94~km~s$^{-1}$ at $z_{\rm simul}=3.5$. Right panel: Comparison to B21 with a pseudo-NB width of 56~km~s$^{-1}$ at $z_{\rm simul}=3$.}
\label{fig:SBsimulation}
\end{figure*}

These fractions, and the associated radii, are contingent on the uncertainties in the Ly$\alpha$ LF, especially its faint-end,  which we here encapsulate by two parameters, the slope  $\alpha_{\rm{LF}}$ and the low-luminosity cutoff $L_{\rm{min}}$. The effects of modifying any of these two are displayed in Fig.~\ref{fig:SBratio} by varying the symbols ($\alpha_{\rm{LF}}$) and colors ($L_{\rm{min}}$). It is evident that varying $L_{\rm{min}}$ has a much weaker effect than adopting a different LF slope. An only slightly shallower LF would reduce the expected SB contribution of faint LAEs drastically to  $\approx20-40$\% at $R\sim 50$~pkpc, whereas a steeper slope would imply that extended Ly$\alpha$ emission could be dominated by discrete objects already from distances of $20-30$~pkpc outwards. It is worth mentioning that if undetected LAEs cluster like the extrapolated HOD models, the minimum luminosity choices of $\log(L_{\rm{min}}/[\rm{erg}\;\rm{s}^{-1}])=37,\;40$ deliver indistinguishable SB ratios as those of Fig.~\ref{fig:SBratio}. If undetected LAEs cluster like the LAEs in MUSE-Wide or MXDF, these ratios increase by (on average) $\approx70$\% and $\approx30$\%, respectively. In Appendix~\ref{appendix:NB}, we show how the SB ratio varies for different pseudo-NB widths.

These are very rough estimates. A sharp cut of the Ly$\alpha$ LF below a certain Ly$\alpha$ luminosity is of course implausible. A more realistic shape would probably involve some smooth turnover towards fainter luminosities. Such subtleties are however beyond the scope of this paper. Clearly more work is needed to study the faint parts of the Ly$\alpha$ luminosity function and its possible (non-)evolution with redshift. Nevertheless, our findings strongly support a scenario in which discrete but individually undetected LAEs are an important component of observed Ly$\alpha$ halos at $R\ga 30-50$~pkpc.

\subsection{Comparison to simulations}
\label{sec:simulations}

Our results are in good agreement with the fundamental prediction of several simulation or modelling studies that faint LAEs in the vicinity of Ly$\alpha$ halos contribute to, and beyond some radii even dominate the observed extended emission \citep[e.g.,][]{lake,ribas16,ribas,mitchell,byrohl}. Nevertheless, the   models and also the predicted Ly$\alpha$ SB profiles due to  "satellite" LAEs vary significantly between different studies.
It is also worth mentioning that some  of our undetected LAEs are, in principle, at the DMH center. 

In the following we compare our Ly$\alpha$ SB profiles calculated from clustering with simulation studies that separate the satellite SB contribution from other powering mechanisms. These are the main differences:

\begin{enumerate}
\item \cite{lake}, hereafter L15, used an adaptive mesh refinement hydrodynamical simulation of galaxy formation \citep{bryan, joung} and modeled most potential sources of Ly$\alpha$ photons and powering mechanisms i.e., star formation from central and satellite galaxies, photon scattering, fluorescence from the UV background, supernova feedback (outflows), and gravitational cooling at $z_{\rm simul}=3.1$. The Ly$\alpha$ emission was modeled using the Monte Carlo radiative transfer code of \cite{zhengmiralda}. The dark matter particle mass is 1.3x$10^7$~M$_\odot$, the simulation box size is 120~$h^{-1}$Mpc, and the spatial resolution $\approx111$~pc. 
Although the derived total Ly$\alpha$ SB profiles matched those measured in \cite{matsuda}, \cite{xue} found that the simulated profile that included star formation from satellites considerably overpredicted the measured SB curves. To extract the SB profile, they employed a projection depth of 224~pkpc to include all scattered photons around the LAEs. 
%caveats:
%1. no agn

%\item \citet[][hereafter MRD16]{ribas16} used analytic recipes to model the CGM (decelerated outflow model from \citealt{DijkstraKramer}, hereafter MRD16a, and clumpy outflow model from \citealt{steidel10}, hereafter MRD16b) and fitting functions to hydrodynamical simulations (EAGLE simulations; \citealt{rahmati}, hereafter MRD16c). They made assumptions for the extension of LAHs based on Monte Carlo simulations of radiative transfer for Ly$\alpha$ photons scattering through the CGM. They further modeled the photoionization rate from central and  satellite galaxies, from which they computed the corresponding fluorescent radiation. For satellites, they only considered the fluorescence SB contribution to the Ly$\alpha$ halos. None of these models was able to reproduce the Ly$\alpha$ SB profiles measured in \cite{matsuda}.
%While \citet[][hereafter MRD16a]{DijkstraKramer} assume that Ly$\alpha$ photons are only produced by central star formation, and then propagate through a clumpy CGM. They thus ignore the off-center star formation and inflow of cold gas.

%caveats:
%1. MRD16a and MRD16b were designed to reproduce observations of (the more massive) LBGs, and might therefore overpredict the surface brightness.
%2. For the central LAE we assume a SFR=10 
%3. clustering from ouchi10
%4. Lya fesc 100\%.

\item \cite{mitchell}, hereafter M21, employed an adaptive mesh refinement zoom-in cosmological radiation hydrodynamics simulation of a single galaxy using the \texttt{RAMSES-RT} code \citep{rosdahl13,rosdahl15}. The stellar and DMH masses are $M_{\star}=10^{9.5}\;\rm{M}_{\odot}$ and $M_h=10^{11.1}\;\rm{M}_{\odot}$, respectively. They modeled the same Ly$\alpha$ powering mechanisms at $3<z<4$ as L15. They also performed Monte Carlo radiative transfer to assess the Ly$\alpha$ emission, tracing the emergent radiation along several different lines of sight and at different cosmic times. The simulation box is a sphere of 150~kpc radius and achieved a spatial resolution of 14~pc, with a characteristic dark matter particle mass of $10^4$~M$_{\odot}$. The simulated total Ly$\alpha$ SB profile matched the one measured in \cite{wisotzki18}, for which they used a projection depth of $\pm150$~pkpc  at $z_{\rm simul}=3.5$ around the LAE.
%caveats:
%1. simulation missing an important component of neutral outflowing gas and dont reproduce the apparent Lya spectral morphology that is redshifted with respect to the systemic velocity
%2. we do not account for photo-heating and ionization by any local active galactic nuclei

\item \cite{byrohl}, hereafter B21, utilized the outcome of the TNG50 \citep{nelson19b, pillepich19} from the IllustrisTNG simulations \citep{pillepich18b}, coupled to their Ly$\alpha$ full radiative transfer code \texttt{VOROILTIS}. They modeled the same Ly$\alpha$ emission sources and powering mechanisms at $2<z<5$ as L15 and M21 and further included a treatment for active galactic nuclei. TNG50 has a box size of 50~Mpc, attains a spatial resolution of $\sim$~100~pc and a dark matter particle mass of 4.5x$10^5$~M$_{\odot}$. This work was able to reproduce the Ly$\alpha$ SB profiles measured in \cite{leclerq}, for which they included all scattered photons from within $\pm100$~pkpc along the line of sight at $z_{\rm simul}=3$ around the LAE.  %, for which they used (last pg of byrohl21)

%caveats:
%1. we do not include the destruction of Lya photons by dust
\end{enumerate}

Although most simulations deliver similar conclusions and their total Ly$\alpha$ SB profiles match various observations, the predicted contribution from satellites to the observed radial profiles at various scales  varies significantly. We compare in Fig.~\ref{fig:SBsimulation} our clustering-based Ly$\alpha$ SB profiles (with $\log(L_{\rm{min}}/[\rm{erg}\; \rm{s}^{-1}])=38.5$, but considering all clustering scenarios) to simulated satellite radial profiles. For this purpose, we recalculate our enhancement factors and SB profiles in order to approximately match our selected pseudo-NB width (i.e., $Z_{\rm NB}$ in Eq.~\ref{eq:boosting}) to the projection depth applied in the simulations. Note that while the pseudo-NB width in observed spectral stacks cannot be narrower than the common linewidth, the typical bandwidths formally converted to real space depths are much larger than simulation boxes. To enable at least a rough comparison we use the simulated depths, which correspond to velocity intervals of $\approx60-100$~km~s$^{-1}$.
We also multiply the simulated profiles by  $(1 +z_{\rm simul} )^4/(1 + \langle z\rangle)^4$ to account for
surface brightness dimming. The simulated contribution is due to the Ly$\alpha$ emission from satellites in L15 (dashed magenta) and M21 (dashed light blue), and to this plus a genuinely diffuse emission powered by these "other halos" in B21 (dashed light gray).%, and to their fluorescence in MRD16a, MRD16b, and MRD16c (gray). %We multiply the simulated profiles by $(1 + z_{\rm pair})^4/(1 + z_{\rm{sim.}})^4$ to account for surface brightness dimming, with $z_{\rm pair}=3.8$ and $z_{\rm{sim.}}$ the redshift of each simulation study. %See pg9 in SB profiles from HETDEX

The profile from L15 is clearly flatter than those from M21 and  B21 and is in this respect similar to our clustering-based SB profiles. This also agrees with the generally observed trend that the radial profiles tend to flatten at larger radii \citep{matsuda, momose14, wisotzki18, hetdex22}. On the other hand,  the SB level predicted by L15 is much higher than ours, already by an order of magnitude than our two upper limit scenarios. The SB levels predicted by M21 and B21 are in better agreement with our upper limit SB estimates, although still higher than our extrapolated "low clustering" scenarios. This may reflect the fact that simulated galaxies in L15 present higher stellar and DMH masses ($M_{\star}=2.9\cdot10^{10}\;\rm{M}_{\odot}$ and $M_h=10^{11.5}\;\rm{M}_{\odot}$) than those of MUSE LAEs or those considered in M21 and B21. At small scales, the M21 and B21 simulated radial profiles are slightly steeper than the high clustering scenario curves of MUSE-Wide and MXDF.

The differences between our clustering-based and the simulated SB profiles could be due to several factors. An overprediction of the simulated number of satellites would naturally lead to an overestimated contribution of discrete LAEs to the Ly$\alpha$ halos. This issue seems to be common to simulations and semi-analytical models, including the IllustrisTNG project and EAGLE simulations %also subhalo abundance matching (SHAM), smoothed particle hydrodynamics (SPH)
\citep[e.g.,][]{okamoto,simha,geha,shuntov}.  
In this context it is interesting to note that our upper limit clustering scenarios give results closer to the simulations than the extrapolations deemed to be more realistic.

There could also be a problem of small number statistics, since  the models by 
%MRD16a and MRD16b focused on the fluorescence contribution from satellites rather than on their genuine Ly$\alpha$ emission \citep{ribas}. Moreover, they were designed to reproduce observations of more massive Lyman Break Galaxies (LBGs). Hence, we should not necessarily expect to obtain the same SB values. Furthermore, MRD16a did not introduce Ly$\alpha$ photon scattering, which would redistribute the emission in the Ly$\alpha$ halos and flatten the SB profiles. In contrast, 
M21 and L15 are based on only one and nine galaxies, respectively. This is probably not an issue for B21, who on the other hand did not account for Ly$\alpha$ destruction by dust, leading to overestimated  luminosities,  possibly also influencing the resulting LAH shapes.

It is of course also possible that the differences are partially driven by our assumptions. Our estimated fraction of undetected LAEs might be lower than the actual one, either because of a steeper luminosity function or because of a lower faint-end cutoff.

\subsection{How to measure the faint LAE SB contribution}
\label{sec:speculation}

While it seems safe to state that a significant contribution of 
faint LAEs to the apparent Ly$\alpha$ SB profiles must be expected, in at least  qualitative agreement between our empirical framework and the predictions by cosmological simulations, it would be very desirable to further constrain their  contribution directly from observations. Here we embark on a few (partly speculative) considerations of how that might be achieved in the future.

%Observational evidence for the presence of undetected satellites is extracted, for instance, from clumpy, asymmetric Ly$\alpha$ halos, halo offsets from the host galaxy, and flattenings in the large scales of the Ly$\alpha$ SB profiles. While \cite{momose16} and \cite{leclerq} did not encounter any of these signs, possibly due to low spatial resolution, \cite{adelaide} found noteworthy offsets between the Ly$\alpha$ and the UV peak emissions and \cite{hetdex22} measured a flattening in the SB profiles at large radii ($R>100$~pkpc). These offsets can be, however, also explained by other mechanisms (i.e., variations in the H{\sc i} column density e.g., due to feedback, outflows or specific H{\sc i} geometries at the scales of the CGM) and the flattening via other processes (i.e., fluorescence from the ultraviolet background, gas cooling or H{\sc i} clouds).

One avenue of investigation could be a comparison between the behaviour of individual and stacked Ly$\alpha$ SB maps at large scales. In principle this should contain essential information about the faint LAE SB contribution: Stacking at random angles invariably contains some degree of azimuthal averaging which smears out the contributions of individual neighboring LAEs, while individual LAEs have neighbours only in certain directions. Using the deepest existing data it might be possible to estimate the incidence of clear asymmetries or other external disturbances in individual Ly$\alpha$ halos and use this to constrain the frequency of marginally detectable close neighbours. 

Another approach could be to compare the outcomes of different stacking methods, since these differ in the sensitivity to asymmetries in the outskirts: A profile derived from a median-stack of Ly$\alpha$ images should show a different contribution from faint external LAEs than a profile obtained as the mean of several azimuthally averaged individual profiles.

Also relevant is the bandwidth chosen to extract the LAE pseudo-NB images from the original IFU data, as this influences the contrast between the "background" contribution of unrelated and, with respect to the central galaxy, unclustered LAEs and the enhancement due to physical neighbor LAEs at similar redshifts. Varying this bandwidth will provide insights about the relative balance between these two.

Observations seem to show that on average, LAH profiles are remarkably self-similar over a wide range of  Ly$\alpha$ luminosities. Yet, if at some radius the contribution from external LAEs prevails, then the self-similarity should break down around that radius, and profiles obtained at different luminosity levels should converge towards the same behaviour. 

At fixed limiting sensitivity, the contribution of faint LAEs to the combined LAH profiles should also decrease rapidly with redshift, simply because of $(1+z)^4$ dimming together with a (roughly) unevolving luminosity function. While \cite{leclerq} did not find any significant Ly$\alpha$ halo size evolution with redshift, their sizes clearly refer to the "inner" LAHs and leave the outer regions still unconstrained. A slightly stronger boundary condition is the similarity of stacked Ly$\alpha$ profiles at $z\approx 3.5$, 4.5, and 5.5 in \cite{wisotzki18}, which could suggest that at least at high redshifts there is indeed a genuinely diffuse component also in outer regions of LAHs. On the other hand, the same phenomenon could arise from a steepening Ly$\alpha$ LF, and/or an increasing clustering strength of LAEs towards higher redshifts, which would then counteract the cosmological dimming. 

We can estimate whether this scenario is actually realistic: While \cite{ouchi17}, HA21, HA23 did not find a significant clustering dependence on redshift, \cite{durkalek} and \cite{khostovan} found a clear increase of clustering strength with cosmic time. To calculate the increase of clustering strength needed to counteract the SB dimming (i.e., $(1+z)^{-4}$) in redshift bins of $\Delta z = 1$, we would need a factor $\approx1.25$ increase in the large-scale bias factor, from $b\approx2.65$ for $3<z<4$ to $b\approx3.30$ for $4<z<5$ and $b\approx4.15$ for $5<z<6$, in broad agreement with the clustering growth found in \cite{durkalek} and \cite{khostovan}.
%from fig3 in khostovan19: at z=3 they find r0=4, z=4 and r0=6, z=5 and r0=6-7, z=6 and r0=12
%from fig8 in durcalek15: at z=1 they find b=1, z=2 and b=2, z=3 and b=3, z=4 and b=4

Ultimately, all these suggested analyses and experiments will require more and better data than currently available. Large LAE samples such as that from the Hobby-Eberly Telescope Dark Energy Experiment (HETDEX; \citealt{hetdex}) or future LAE integral field spectrographs such as BlueMUSE \citep{richard} may deliver the data quantity and quality to discriminate between genuinely diffuse Ly$\alpha$ emission in the circum- and intergalactic medium on the one hand, and "fake diffuse" contributions from faint undetected LAEs on the other hand.

%%%%%%%%%%%%%%%%%%%%%%%%%%%%%%%%%%%%%%%%%%%%%%%%%%%%%%%%%%%%%%%%%%%%%%%%%%%%%%%%%%%%%%%%%%%%%%%%%%%%%%%%%%%%

\section{Conclusions}
\label{sec:conclusions}
In this paper we turn the observed clustering properties of a sample of 1265 Ly$\alpha$ emitting
galaxies (LAEs) at $3 < z < 5$ from the MUSE-Wide survey  into implications for the spatially extended Ly$\alpha$ emission in the circumgalactic medium (CGM) of LAEs. All sources have spectroscopic redshifts and their median Ly$\alpha$ luminosity is $\log (L_{{\rm Ly}\alpha}/[\rm{erg\;s}^{-1}])\approx 42.27$.

We use  halo occupation distribution (HOD) modeling to represent the clustering of our LAE set. Because the extended Ly$\alpha$ emission around LAEs is commonly measured using (pseudo-)narrow-band (NB) filters centered on the Ly$\alpha$ wavelength, we include redshift-space distortions both at large (Kaiser infall) and  small (Finger of God effect) scales. %We find that DMHs typically host one (central) LAE, and satellite LAEs are rarely present in our data, supporting the results derived in HA23.

We then extrapolate the HOD statistics inwards towards smaller radii and combine them with assumptions about the Ly$\alpha$ emitter luminosity function (LF). We only consider the emission from undetected LAEs, less luminous than those from our current dataset. Together, these two ingredients are  transformed into a (upper limit) Ly$\alpha$ surface brightness (SB) profile, which belongs to individually undetected close neighbors that cluster like those in the MUSE-Wide sample. We derive a maximum SB values of $\rm{SB}\approx2.5\cdot10^{-20}$~erg~s$^{-1}$~cm$^{-2}$~arcsec$^{-2}$, assuming a luminosity of the undetected LAEs of $\log (L_{{\rm Ly}\alpha}/[\rm{erg\;s}^{-1}])=38.5$.

We consider various alternative clustering scenarios for the undetected sources. We first assume that these follow the same clustering properties as the LAEs in the MUSE-Extremely Deep Field (MXDF; one order of magnitude fainter than those in MUSE-Wide but still more luminous than undetected LAEs) and compute the corresponding (upper limit) radial profile ($\rm{SB}\approx2\cdot10^{-20}$~cgs). We then extrapolate the clustering properties of MUSE-Wide and MXDF LAEs down to $\log(L_{\rm{Ly}\alpha}/[\rm{erg}\; \rm{s}^{-1}])\approx37.0,\;38.5,\;40.0$ and use the resulting HOD models to estimate the actual  Ly$\alpha$ SB profiles from undetected LAEs. We obtain a maximum SB of $\rm{SB}\approx4\cdot10^{-21}$~cgs.

We compare our $\log (L_{{\rm Ly}\alpha }/[\rm{erg}\;\rm{s}^{-1}])=38.5$ undetected Ly$\alpha$ SB profile to the LAE stacking experiment performed in \cite{wisotzki18} to address the question  whether undetected LAEs play a pivotal role in the formation of the extended Ly$\alpha$ halos.  Assuming a simple Schechter LF with a reasonable intermediate faint-end slope ($-1.94\leq\alpha_{\rm{LF}}\leq-1.84$) and a lower limit for Ly$\alpha$ luminosities of the undetected LAEs ($\log(L_{\rm{Ly}\alpha}/[\rm{erg}\; \rm{s}^{-1}])\approx38.5$), we find that the stellar irradiation from those undetected LAEs can dominate the excess surface brightness at large scales ($R\gtrsim 50$~pkpc). On the other hand, the  Ly$\alpha$ SB profile at small scales ($R\lesssim 20$~pkpc) cannot be explained by undetected sources  and may be better  explained by a genuinely diffuse origin. 
More luminous LAEs ($\log(L_{\rm{Ly}\alpha}/[\rm{erg}\; \rm{s}^{-1}])\approx40$) reproduce at best 4\%, 40\%, and 100\% of the extended emission at $R=15,\;40,\;65$~pkpc, respectively.

We also compare our estimated Ly$\alpha$ SB profiles  with simulation studies. Although we agree that faint LAEs dominate the SB of the Ly$\alpha$ halos at large scales, the shape of the radial profiles and the contribution to the total Ly$\alpha$ SB profiles differ.  
%when does satellite contribution dominate over other factors?
%MRD16 and L15 around 30~pkpc
%B21 we find that the flattening of LAH profiles at large radii becomes dominated by photons originating from other nearby halos rather than diffuse emission itself. 
%dominates around 50~pkpc
%M21 around 20~pkpc
While the simulated faint LAE SB profiles generally decrease rapidly with distance, our derived radial profiles have shallow slopes, likely leading to the flattening at $R\gtrsim30$~pkpc seen in observed Ly$\alpha$ SB profiles  Overall, most simulated profiles, together with our estimations, infer a faint LAE contribution  of the same order of magnitude. %This also agrees with \cite{mitchell} but it is somewhat lower than the calculations form \cite{lake,DijkstraKramer,rahmati}.

Although these faint LAEs are likely the most significant source of luminosity for the outer parts of observed Ly$\alpha$ halos, beyond the scales probed by most observations of single objects ($R>50$~pkpc), the actual point at which this contribution starts to become important depends crucially on the shape of the Ly$\alpha$ luminosity function, in particular its faint-end. We also suggest a few experiments to directly constrain the faint LAE SB contribution from observations.

%-----------------------------------------------------------
\begin{acknowledgements}
      The authors give thanks to the staff at ESO for extensive support during the visitor-mode campaigns at Paranal Observatory. We thank the eScience group at AIP for help with the functionality of the MUSE-Wide data release webpage. T.M. thanks
      for financial support by CONACyT Grant Cient\'ifica B\'asica \#252531 and by UNAM-DGAPA (PASPA, PAPIIT IN111319 and IN114423). L.W. and J.P. by the Deutsche Forschungsgemeinschaft through grant Wi 1369/31-1. The data were obtained with the European Southern Observatory Very Large Telescope, Paranal, Chile, under Large Program 1101.A-0127. This research made use of Astropy, a community-developed core Python package for Astronomy \citep{astropy}. We also thank the referee for an useful and constructive report.
\end{acknowledgements}

%-------------------------------------------------------------------
\clearpage

\begin{appendices}

\section{Clustering comparison from LAE subsets of the MUSE-Wide survey}
\label{app:same_clustering}

In this work, we have utilized the clustering constraints derived in  HA23 for a subset of 1030 MUSE-Wide LAEs at $3<z<6$ to estimate the contribution of undetected LAEs to the apparent Ly$\alpha$ halos observed at $3<z<5$. To be consistent with the Ly$\alpha$ halo measurements, we focused on a subsample of 1265 LAEs at $3<z<5$ from the MUSE-Wide survey. In the following, we demonstrate that the LAEs in  these two datasets have nearly identical clustering strengths.

In Fig.~\ref{fig:old-new}, we represent in blue the K-estimator measurements obtained in HA23 for the $3<z<6$ LAE subsample and in red the corresponding measurement for the $3<z<5$ LAE subset of this work. The clustering strengths are in excellent agreement. The clustering uncertainties for the former sample are (on average) 2\%
larger than for the latter dataset.
The best-fit HOD model and, thus, the large-scale bias factor and typical DMH masses are indistinguishable.

\renewcommand\theequation{B.\arabic{equation}}
\setcounter{equation}{0}
\section{Clustering enhancement factor derivation}
\label{app:boosting}
The surface brightness (SB) at cosmological distances is defined as
\begin{equation}
\label{eq:noclusteringSB}
     {\rm SB} =\int_{Z}\epsilon(Z)\frac{[D_{\rm{A}}(z)(1+z)]^2}{4\pi D_{\rm{L}}(z)^2}{\rm d}Z\approx \frac{\epsilon\Delta Z}{4\pi(1+z)^2},
\end{equation}
where $\epsilon(Z)$ is the comoving volume emissivity as a function of radial comoving separation, $Z$, and $D_{\rm{L}}$, $D_{\rm{A}}=D_{\rm{L}}/(1+z)^2$ are the luminosity and angular size distances at redshift $z$, respectively. The radial comoving separation $Z$ corresponds to the (pseudo-)NB width
employed to stack LAE images in the measurement of Ly$\alpha$ SB profiles. This definition considers the shape and expansion history of the universe and assumes that objects are randomly distributed. 

To account for the clustering of galaxies, we include the excess probability d$P$ of finding a galaxy $i$ in a volume element d$V$ at a separation $R=\sqrt{R_{ij}^2+Z_{ij}^2}$ from another galaxy $j$, i.e.,  the two-point correlation function \citep[$\xi(R)$; ][]{peebles1980}
\begin{equation}
    {\rm d}P=n\cdot[1+\xi(R)]\cdot{\rm d}V \propto [1+\xi(R)]\cdot\frac{{\rm d}Z}{{\rm d}z}{\rm d}z,
\end{equation}
where $n$ is the mean number density of the galaxy sample and $R_{ij}$ is the transverse separation between the galaxy pair. The two-point correlation function of interest is, in fact, a cross-correlation function between detected and undetected LAEs. This is because  Ly$\alpha$ SB profiles represent SB as function of distance from the central LAE outwards, not as function of  observed LAE$-$observed LAE separation. 
Hence, the SB that accounts for galaxy clustering is 
\begin{multline}
\label{eq:clusteringSB}
    {\rm SB} =\int_{Z}\frac{\epsilon(Z)}{4\pi (1+z)^2}\cdot[1+\xi(R)]\cdot\frac{{\rm d}Z}{{\rm d}z}{\rm d}z\approx \\
    \frac{\epsilon}{4\pi(1+z)^2}\int_{Z}[1+\xi(R)]\cdot{\rm d}Z.
\end{multline}

We derive the clustering enhancement factor $\zeta(R)$ from the comparison between Eqs.~\ref{eq:noclusteringSB} and \ref{eq:clusteringSB},
\begin{equation}
    \zeta(R)= 1+\int_{Z}\xi(R)\cdot\frac{{\rm d}Z}{\Delta Z}.
\end{equation}

\renewcommand\thefigure{\thesection.\arabic{figure}}
\setcounter{figure}{0}
\renewcommand\thetable{\thesection.\arabic{table}}
\setcounter{table}{0}
\renewcommand\theequation{A.\arabic{equation}}
\setcounter{equation}{0}
\begin{figure}[tb]
\centering
\includegraphics[width=\columnwidth]{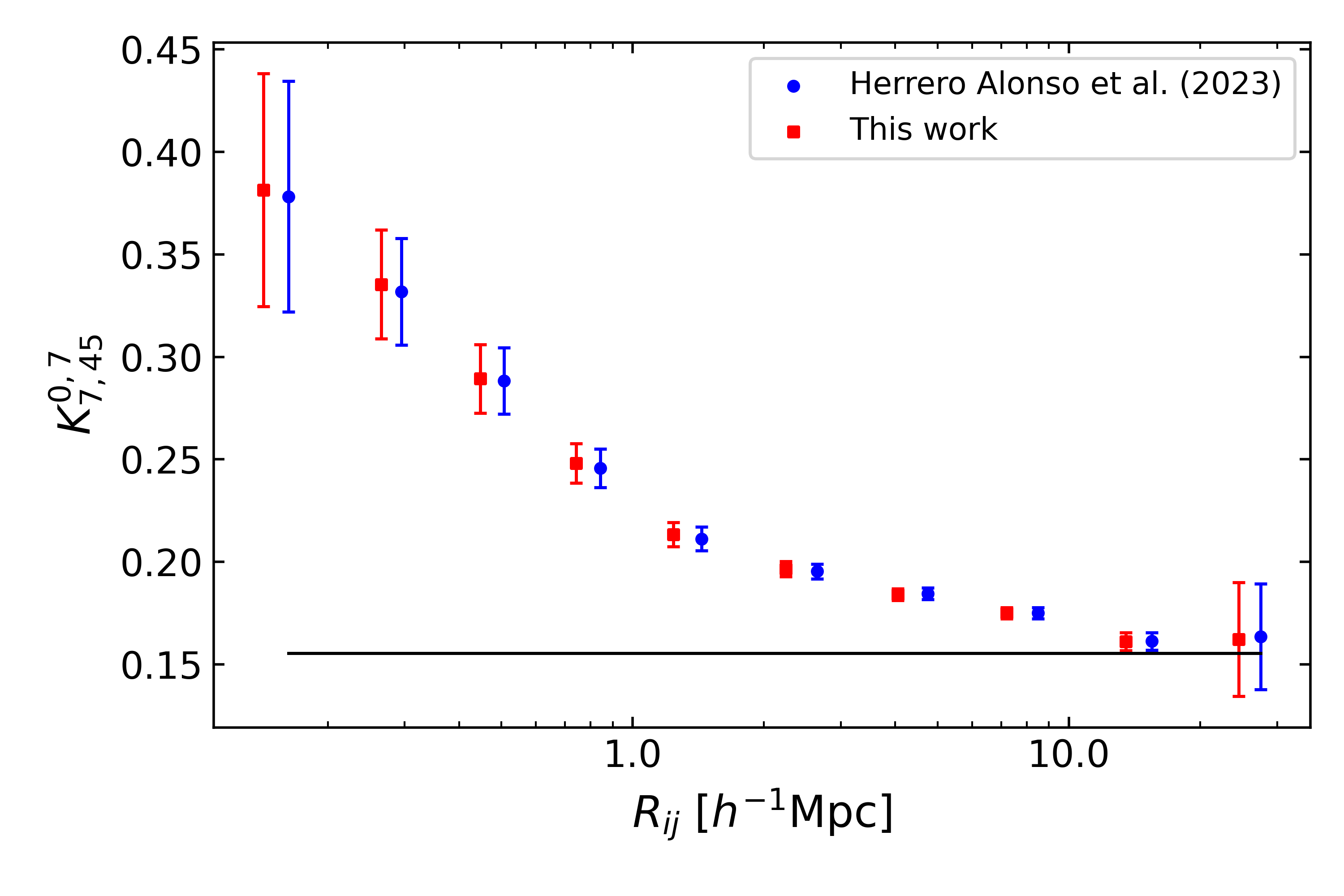}
\caption{Clustering strength as measured by the K-estimator \citep{adelberger} for the LAE dataset considered in HA23 at $3<z<6$ (blue) and that of the LAE sample at $3<z<5$ used in this work (red). The black baseline represents the expected value for  an unclustered sample. The error bars are Poissonian. The red measurements have been shifted along the x-axis for visual purposes.}
\label{fig:old-new}
\end{figure}

\renewcommand\thefigure{C.\arabic{figure}}
\setcounter{figure}{0}
\setcounter{equation}{0}

\section{Effect of the velocity bandwidth on the faint LAE contribution to the extended LAHs}
\label{appendix:NB}

To compute the faint LAE SB profiles (colored lines in Fig.~\ref{fig:SB_clustering})  and to modify the stacking analysis of \cite{wisotzki18} (data points in Fig.~\ref{fig:SB_clustering}),
we adopted a fixed velocity bandwidth of 600~km~s$^{-1}$.
The selected pseudo-NB width  influences the Ly$\alpha$ SB values of the stacked radial profile,  the background Ly$\alpha$ SB level due to discrete faint LAEs, and the clustering enhancement factor (Eq.~\ref{eq:boosting}). 
\begin{figure*}[tb]
\centering
\begin{tabular}{c c}
  \centering
  \includegraphics[width=.47\linewidth]{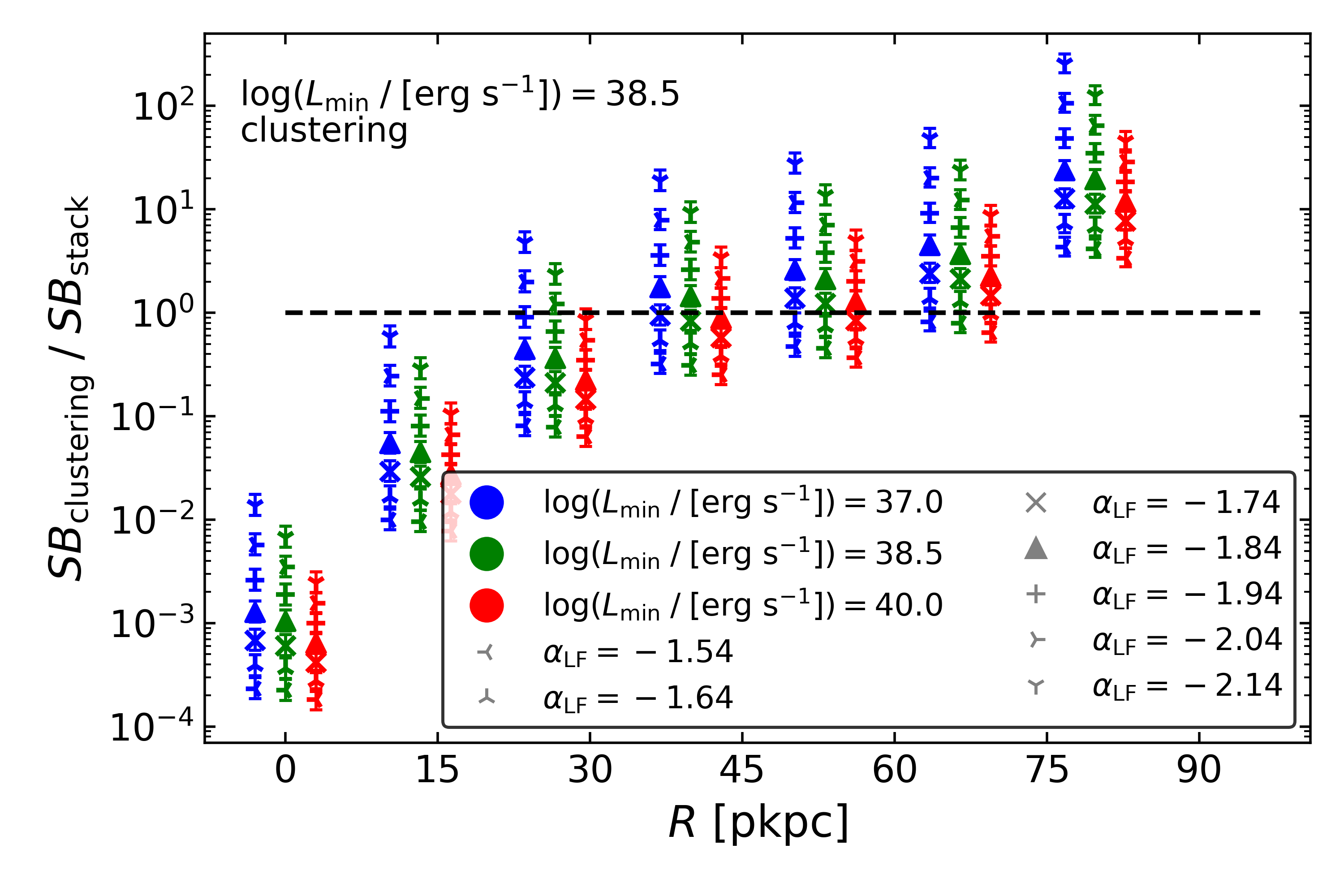}
\end{tabular}
\begin{tabular}{c c}
  \centering
  \includegraphics[width=.47\linewidth]{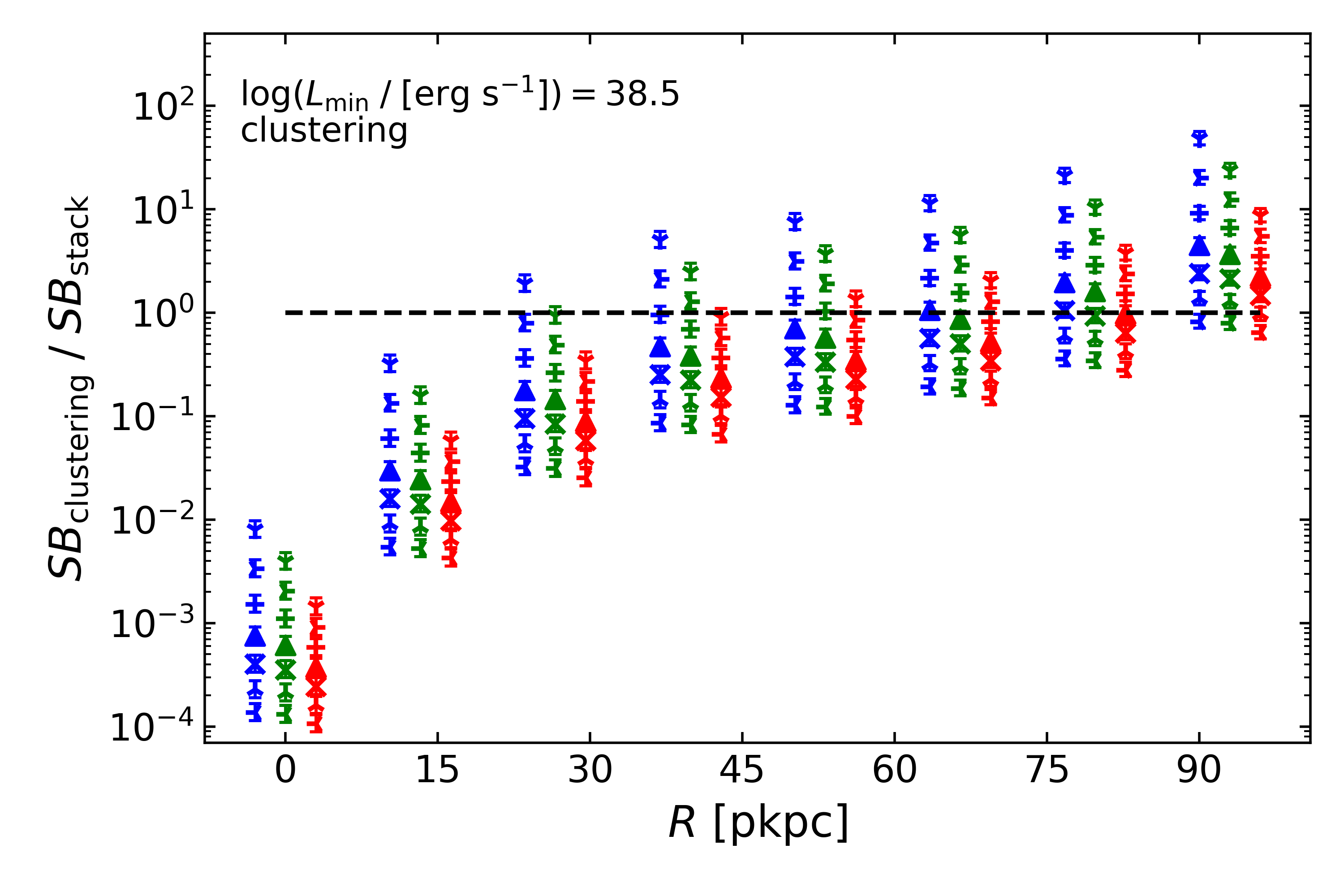}
\end{tabular}
\caption{SB ratio of the Ly$\alpha$ SB profile from undetected LAEs and the modified stacked profile from \cite{wisotzki18}. Left: Same as Fig.~\ref{fig:SBratio} but for a pseudo-NB width of 400~km~s$^{-1}$. Right: Same for a pseudo-NB width of 800~km~s$^{-1}$.} 
\label{fig:SBratioNBs}
\end{figure*}
The general trend is that, for broader  velocity widths, the stacked SB profile notably rises in the outer regions ($R>20$~pkpc), the background SB slightly increases, and the clustering enhancement factor  significantly declines (see Fig.~\ref{fig:boosting}). The interplay between these three factors is thus reflected in the SB ratio of Fig.~\ref{fig:SBratio}.

In Fig.~\ref{fig:SBratioNBs}, we recompute the ratio between the faint LAE SB profile and the stacked  one for a pseudo-NB width of 400~km~s$^{-1}$ (left panel) and 800~km~s$^{-1}$ (right panel). For broader bandwidths, the SB ratios decrease  because the corresponding decline of the clustering enhancement factor exceeds the smaller rises in the stacked radial profile and  the background SB level. While a broadening of the pseudo-NB width from 400~km~s$^{-1}$ to 600~km~s$^{-1}$ causes a maximum 40\% drop of the SB ratios, an increase from 600~km~s$^{-1}$ to 800~km~s$^{-1}$ further decreases the SB ratios in at most 20\%. Note that for a velocity offset of 600~km~s$^{-1}$ or 800~km~s$^{-1}$ the stacked Ly$\alpha$ SB value at $R=90$~pkpc is an upper limit (see e.g., Fig.~\ref{fig:SB_clustering}). For 400~km~s$^{-1}$, on the other hand, there is no detection.

\end{appendices}

\end{document}